\documentclass[prd,nofootinbib,twocolumn]{revtex4}
\usepackage{graphicx}
\usepackage{bm}
\usepackage{float}
\usepackage{subfigure}
\usepackage{amsmath}
\usepackage{amsfonts}
\usepackage{color}
\usepackage{slashed}
\usepackage{fancyhdr}
\def\be{\begin{equation}}
\def\ee{\end{equation}}
\def\bed{\begin{description}}
\def\eed{\end{description}}

\def\bea{\begin{eqnarray}}
\def\eea{\end{eqnarray}}

\def\ba{\begin{array}}
\def\ea{\end{array}}

\def\u1{$U(1)$}
\def\suu1{$SU(2)\times U(1)$}

\begin{document}

\title{Aharonov-Bohm Radiation of Fermions}

\author{Yi-Zen Chu$^1$, Harsh Mathur$^1$ and Tanmay Vachaspati$^{1,2}$}
\affiliation{
$^1$CERCA, Department of Physics, Case Western Reserve University,
10900 Euclid Avenue, Cleveland, OH 44106-7079, \\
$^2$Institute for Advanced Study, Princeton, NJ 08540}
%\date{}

\begin{abstract}
\noindent
We analyze Aharonov-Bohm radiation of charged fermions from
oscillating solenoids and cosmic strings. We find that the
angular pattern of the radiation has features that differ
significantly from that for bosons. For example,
fermionic radiation in the lowest harmonic is approximately
isotropically distributed around an oscillating solenoid, whereas
for bosons the radiation is dipolar. We also investigate the
spin polarization of the emitted fermion-antifermion pair. Fermionic
radiation from kinks and cusps on cosmic strings is shown to depend
linearly on the ultraviolet cut-off, suggesting strong emission at
an energy scale comparable to the string energy scale.
\end{abstract}

\maketitle

\section{Introduction}

The Aharonov-Bohm (AB) interaction \cite{Aharonov:1959fk,AlfordWilczek} between
charged particles and thin magnetic fluxes is of much interest as
it provides a physical consequence of a pure gauge field with
vanishing field strength but non-trivial topology. Further, the
physical effects emerge only in quantum theory and hence the AB
interaction provides an example of a quantum, topological interaction.

The classic Aharonov-Bohm setup involves scattering an electron
off a solenoid, with non-trivial scattering obtained even for an
arbitrarily thin solenoid, whereby the electron is exclusively
localized in a region of vanishing magnetic field. A novel
feature of the scattering is a periodic dependence of the
scattering cross-section on the magnetic flux through the solenoid.
If $\Phi$ denotes the magnetic flux in the solenoid and
$e$ the electron charge, the cross-section is proportional
to $\sin^2(\pi\epsilon)$ where $\epsilon \equiv e\Phi/2\pi$.

The classic AB setup was recently extended in another direction
\cite{AB_Bosons}, where it was shown that an oscillating solenoid
in vacuum can produce charged particle-antiparticle bosons from
the vacuum due to the AB interaction. The AB radiation rate also
has the characteristic $\sin^2(\pi \epsilon)$ dependence on the
magnetic flux.

AB radiation is relevant to the evolution of cosmic strings, which
are similar to solenoids, except the magnetic flux within
them is massive, unlike electromagnetic fluxes in laboratory
solenoids. Moreover, the gravitational analog of the AB effect can
cause cosmic strings to emit light, even if the fields composing the
cosmic string are unrelated to electromagnetic
fields \cite{Garriga:1989bx,AB_Bosons}.

In this paper we will investigate {\it fermionic} AB radiation.
One motivation is that the electron is a fermion. Hence
fermionic AB radiation is what is relevant to oscillating
solenoids. The investigation is also relevant, for instance, to neutrino
emission from cosmic strings by the AB process. A second motivation
is that the spin of the fermion adds another degree of freedom
to the emission and the polarization properties of the radiation
are of interest.

Our results show a significant difference between AB radiation
of bosons and fermions. For example, if a solenoid aligned with
the $z$-axis oscillates along the $x$-direction, the angular
distribution of bosonic AB radiation is peaked in the $y$-direction.
Fermionic AB radiation, however, is (approximately) isotropically
distributed.

The outline of the paper is as follows.
In Secs.~\ref{abfermionpairproduction} through
\ref{cosmicstringloops}, we use conventional interaction picture
perturbation theory to calculate the fermion-antifermion pair
production rate in the small AB phase ($\epsilon$) limit. We
consider AB radiation from an infinite straight solenoid
oscillating perpendicular to its length in
Sec.~\ref{infinitestraightstring}, a cosmic string loop with
kinks in Sec.~\ref{kinkyloops}, and a cosmic string loop with
cusps in Sec.~\ref{cuspyloops}. In
Sec.~\ref{movingframesperturbationtheory}, we
solve the problem using a different technique that does not
assume that $\epsilon$ is small. We call this the ``moving
frames'' scheme and use it to obtain the $\sin(\pi\epsilon)$
dependence of the radiation on the AB phase, provided the
motion of the solenoid is slow. We conclude in
Sec.~\ref{conclusions} and describe our conventions in
Appendix~\ref{conventions}.

\section{AB Fermion Pair Production}
\label{abfermionpairproduction}

\subsection{Setup}
\label{setup}

The interaction of fermions with the gauge potential of the
thin solenoid or string is contained within the Dirac action
\begin{equation}
\label{diracaction}
S_\psi \equiv \int d^4x \ \bar{\psi} \left( i\slashed{D} - m \right) \psi.
\end{equation}
(See Appendix~\ref{conventions} for conventions.) The
relevant interaction term is
\begin{equation}
L_{\rm int} = e A_\mu {\bar \psi} \gamma^\mu \psi
\end{equation}
where $A_\mu$ is the classical solution around the flux tube
\cite{AlfordWilczek}
\begin{equation}
A_\nu = \frac{\Phi}{2} \epsilon_{\mu\nu\alpha\beta} \partial^\mu
       \frac{1}{\partial^2} S^{\alpha\beta}
\end{equation}
with
\begin{eqnarray}
S^{\alpha\beta}(x) &=& \int d\tau d\sigma \sqrt{-\gamma}
     \epsilon^{ab} ~ \partial_a  X^\alpha \partial_b X^\beta
                 \delta^{(4)} (x-X(\sigma,\tau))
\nonumber \\
&& \hskip -1 cm = \int d\tau d\sigma
     ( {\dot X}^\alpha {X^\beta} ' - {\dot X}^\beta {X^\alpha} '
)
                 \delta^{(4)} (x-X(\sigma,\tau))
\end{eqnarray}
and $X^\mu (\sigma, \tau)$ gives the position of the flux tube
in terms of world-sheet coordinates $\sigma$ and $\tau$.

We will need the Fourier transform of $A_\mu$ and this
is given by
\begin{equation}
{\tilde A}_\nu = - i \frac{\Phi}{2} \epsilon_{\mu\nu\alpha\beta}
       \frac{k^\mu}{k^2} {\tilde S}^{\alpha\beta}
\end{equation}
with
\begin{eqnarray}
{\tilde S}^{\alpha\beta} (k) &=&
                \int d^4x ~ e^{+ik\cdot x} S^{\alpha\beta}(x)
                \nonumber \\
                     &=& \frac{1}{2} \int dt \int d\sigma
        \frac{\partial X^{[\alpha}}{\partial t}
     \frac{\partial X^{\beta]}}{\partial \sigma} e^{ik_\mu X^\mu}
\label{Smunu}
\end{eqnarray}
where $\tau = t$ and superscripts within square brackets are antisymmetrized.

In the case of a straight solenoid, we will impose the dynamics
by hand and consider oscillatory motion. In the case of a
cosmic string, the dynamics will be given by the Nambu-Goto
action as discussed in Sec.~\ref{cosmicstringloops}.

\subsection{Pair production amplitude}
\label{pairprodamp}

\begin{figure}
\begin{center}
\includegraphics[width=1.25in]{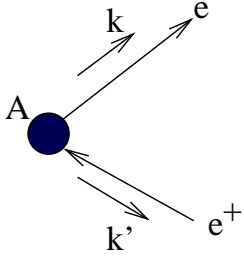}
\end{center}
\caption{Feynman diagram for fermion-antifermion (which we have named
$e^+e^-$ for convenience) pair production from a magnetic flux tube
in motion. The black dot represents the classical gauge field
of the flux tube and the solid lines are the outgoing electron and
positron states.
}
\label{PairProductionDiagram}
\end{figure}

The amplitude for pair production
is then given by the Feynman diagram in Fig.~\ref{PairProductionDiagram}.
\begin{align*}
\mathcal{M}(0 \to e^+e^-) &=
\frac{e\Phi}{p^2} \epsilon_{\mu\nu \alpha\beta} p^\mu
     \mathcal{J}^\nu \widetilde{\mathcal{S}}^{\alpha\beta} \\
\mathcal{J}^\nu(k,s;k',s') &\equiv
\frac{\bar{u}_k^s \gamma^\nu v_{k'}^{s'}}{\sqrt{2 k_0} \sqrt{2 k'_0}}\, ,
\ \  p^\mu = k^\mu + k'^\mu
\end{align*}
where the $s$ and $k$ are the spin and momentum labels for the outgoing
fermion state and the primed labels are for the anti-fermion.

It will turn out, for all three cases of interest in this paper,
$\widetilde{\mathcal{S}}^{\alpha\beta}$ can be factorized into an antisymmetrized
product of two independent integrals,
\begin{align*}
\widetilde{\mathcal{S}}^{\alpha\beta} = \frac{1}{2} I_+^{[\alpha} I_-^{\beta]}
\end{align*}
Furthermore, as can be checked explicitly, the $I_\pm^\mu$ and
electromagnetic current $\mathcal{J}^\mu$ are
conserved\footnote{For $I^\alpha_\pm$, see
Eq.~\eqref{IplusIminus}, and more explicitly,
\eqref{Iplusminus_infinitestring}, \eqref{Iplusminus_kinkyloop} and
\eqref{Iplusminus_cuspyloop}. The identity for $\mathcal{J}^\mu$
follows from the free massive Dirac equations,
$(\slashed{k}'+m)v_{k'}^{s'} = 0={\bar u}^s_k (\slashed{k} -m)$.}
\begin{align*}
p_\mu I^\mu_\pm = p_\mu \mathcal{J}^\mu = 0
\end{align*}
This allows us to re-write
\begin{align*}
\epsilon_{\mu\nu\alpha\beta} p^\mu \mathcal{J}^\nu \widetilde{\mathcal{S}}^{\alpha\beta}
= \frac{p^2}{p_0} \mathcal{\bf J} \cdot ({\bf I}_+ \times {\bf I}_-)
\end{align*}
so that the amplitude now reads
\begin{align}
\label{GeneralAmplitude}
\mathcal{M} =
\frac{2\pi\epsilon}{p_0} \mathcal{\bf J}\cdot ({\bf I}_+ \times {\bf I}_-),
\ \ \epsilon \equiv \frac{e\Phi}{2\pi}
\end{align}

{\it Total rate:} Our normalization of the Dirac spinors
\eqref{planewavenormalization} is such that the square of the
amplitude itself is the total number of fermion-antifermion pairs
produced. First let us evaluate the total rate, and hence
sum over the spins of the final particles. In squaring the amplitude, we may then exploit the spin sums
\begin{align*}
\sum_s u_k^s \bar{u}^s_k = \slashed{k} + m, \qquad
\sum_{s'} v_{k'}^{s'} \bar{v}^{s'}_{k'} = \slashed{k}' - m,
\end{align*}
and the Clifford algebra $\{\gamma^\mu,\gamma^\nu\} = 2\eta^{\mu\nu}$
to deduce,
\begin{align*}
\sum_{s,s'} \mathcal{J}^\mu (\mathcal{J}^\nu)^* &=
\frac{k'^\mu k^\nu + k'^\nu k^\mu - (m^2 + k \cdot k') \eta^{\mu\nu}}
{k_0 k'_0}
\end{align*}
Within the small AB phase approximation, the result of this spin
sum accounts for the entire difference in angular distribution for
the outgoing fermion-antifermion pairs from that of the scalar case
in \cite{AB_Bosons}. We see that, even at the level of unpolarized
rates, the spin of the particles interacting with the vector potential
give rise to significant observational signatures.

For comparison, in the bosonic case the corresponding
quantity is \cite{AB_Bosons}
\begin{align*}
\mathcal{J}^\mu (\mathcal{J}^\nu)^* &=
                    \frac{(k-k')^\mu (k-k')^\nu}{k_0 k'_0}
\end{align*}
The difference in this expression versus the expression
for the fermionic case gives rise to different angular
distributions for the AB radiation of bosons and fermions.
In particular, AB bosonic radiation vanishes when
$k^\mu = {k'}^\mu$ whereas fermionic radiation does not.

Now the square of the amplitude, summed over the possible spins
of the outgoing particles, reads
\begin{align}
\label{GeneralAmplitudeSquared}
\sum_{s,s'} |\mathcal{M}|^2 &=
\frac{(2\pi)^2 \epsilon^2}{p_0^2 k_0 k'_0}
\bigg\{ \left\vert{\bf I}_+ \times {\bf I}_-\right\vert^2
\left( m^2 + k \cdot k' \right) \\
&+ \left( \left({\bf I}_+ \times {\bf I}_-\right) \cdot
{\bf k} \ \left({\bf I}_+ \times {\bf I}_-\right)^* \cdot
{\bf k}' + \text{c.c.} \right) \bigg\} \nonumber
\end{align}
with the ``c.c." representing complex conjugation of the term preceding
it.

The $\left\vert{\bf I}_+ \times {\bf I}_-\right\vert^2$ will contain
an infinite series involving the square of $\delta$-functions, of the
form $(\delta(p_0-\ell\Omega))^2$ where $\Omega$ is the
characteristic frequency of oscillation and $\ell$ is the radiation
harmonic. This may be interpreted as a single $\delta$-function
multiplied by an infinite constant corresponding to the total duration
of time divided by $2\pi$:
\begin{align}
\label{deltafunctionsquared}
(\delta(p_0-\ell\Omega))^2 &\to
\delta(p_0-\ell\Omega) \frac{T}{2\pi}
\end{align}
where $T$ is the total time duration over which the radiation
is calculated.
This can be justified formally by using the integral representation
of the $\delta$-function and setting the exponential to unity,
\begin{align*}
\delta(0) \to
\lim_{T\to \infty} \lim_{\alpha \to 0}
\int_{-T/2}^{T/2} \frac{dt}{2\pi} e^{i\alpha t}
= \frac{T}{2\pi}
\end{align*}
The {\it rate} of pair production is then given by
$\sum_{s,s'} |\mathcal{M}|^2/T$,
integrated over all kinematically possible momenta of the outgoing
particles
\begin{align}
\label{pairproductionrate}
\frac{dN}{dt} &= \frac{(2\pi)^2 \epsilon^2}{T}
  \int \frac{d^3 k}{(2\pi)^3} \int \frac{d^3 k'}{(2\pi)^3}
  \frac{1}{p_0^2 k_0 k'_0} \\
&\times \bigg\{ \left\vert{\bf I}_+ \times {\bf I}_-
\right\vert^2 \left( m^2 + k \cdot k' \right) \nonumber \\
&\qquad + \left( \left({\bf I}_+ \times {\bf I}_-\right) \cdot {\bf k}
\ \left({\bf I}_+ \times {\bf I}_-\right)^* \cdot {\bf k}' +
\text{c.c.} \right) \bigg\} \nonumber
\end{align}

{\it Spin-dependence:}
If one wishes to evaluate the spin-dependence of the scattering it is convenient to work out the amplitudes for the creation of particles and anti-particles with definite helicity. In other words the
particle of momentum $k$ is assumed to have its spin aligned or
anti-aligned with the momentum ${\mathbf k}$ corresponding to positive or negative helicity. This is accomplished by evaluating the current ${\cal J}( k, s; k', s' )$ taking the spinors $u_k^s$ and $v_{k'}^{s'}$ to be of definite helicity (see Appendix~\ref{conventions}).
An example of such a spin-dependent pair production rate is given in
Sec.~\ref{infinitestraightstring}.

The spin-dependence of pair production has a general feature that can be deduced immediately from the form of the matrix element ${\cal M}$.
Let us suppose that
${\mathbf I}_+ \times {\mathbf I}_- \propto \hat{{\mathbf m}}$
where $\hat{{\mathbf m}}$ is a unit vector. This assumption will be
seen to be valid below for the cases of a straight oscillating string
and for degenerate kinky loops discussed below in
Secs.~\ref{infinitestraightstring} and \ref{kinkyloops}. For this
case one can show that
\begin{equation}
{\cal M} \propto {\mathbf J} \cdot \hat{{\mathbf m}} =
\xi_{s}^{\dagger} S_{\Sigma, 0} \xi_{s'} +
\xi_{s}^{\dagger} {\mathbf S}_{\Sigma}
\cdot {\boldsymbol \sigma} \xi_{s'}
\label{eq:spinfeature}
\end{equation}
Here
{\allowdisplaybreaks
\begin{align}
S_{\Sigma, 0} &=
2 i \hat{{\mathbf m}} \cdot ( \hat{{\mathbf n}}' \times
\hat{{\mathbf n}} ) \sinh \frac{\zeta}{2} \sinh \frac{ \zeta'}{2},
\nonumber \\
{\mathbf S}_{\Sigma} &= 2 \cosh \frac{\zeta}{2} \cosh \frac{\zeta'}{2} \hat{{\mathbf m}} \label{eq:ssigma} \\
+& 2 \sinh \frac{\zeta}{2} \sinh
      \frac{\zeta'}{2} \left[
(\hat{{\mathbf n}} \cdot \hat{{\mathbf m}} ) \hat{{\mathbf n'}} +
(\hat{{\mathbf n'}} \cdot \hat{{\mathbf m}} ) \hat{{\mathbf n}}
- (\hat{{\mathbf n}} \cdot \hat{{\mathbf n'}} ) \hat{{\mathbf m}} \right] \nonumber.
\end{align}}
The rapidity is defined via $\cosh \zeta = k^0/m$,
$\cosh \zeta' = k'^0/m$ and $\hat{{\mathbf n}}$
and $\hat{{\mathbf n'}}$ are unit vectors along the directions of
the momenta ${\mathbf k}$ and ${\mathbf k'}$ respectively.
Eq.~(\ref{eq:spinfeature}) is a simple consequence of the forms of
the spinors $u_k^s$ and $v_{k'}^{s'}$ given in Appendix~\ref{conventions}.
Here $\xi_s$ is a spinor that corresponds to the spin of the particle in its rest frame; $\xi_{s'}$ to the spin of the anti-particle in its rest
frame. Thus we see that if the particle is measured to be up along
the direction ${\mathbf S}_{\Sigma}$ in its rest frame, the
anti-particle will definitely be down along the same direction
in its rest frame and vice versa. To see this explicitly, if
$\xi_s^T = (1,0)$ then the amplitude is maximized by taking
$\xi_{s'}^T=(1,0)$. However this choice of spinors means that
the particle is spin up and the antiparticle is spin down. Hence
there is a definite anti-correlation in the spin of the produced
particles.

\section{Infinite, Straight Solenoid}
\label{infinitestraightstring}

In this section we will consider the pair production due to an
infinite, straight solenoid aligned parallel to the $z$-axis,
moving in a sinusoidal fashion along the $x$-axis. Hence,
with $\sigma =z$,
\begin{align}
\label{X_infinitestraightstring}
X^\alpha(t,z) &= \left( t, \xi(t), 0, z \right) \\
\xi(t) &\equiv \frac{v_0}{\Omega} \sin(\Omega t), \ \ \Omega > 0 \nonumber
\end{align}
Putting \eqref{X_infinitestraightstring} into \eqref{Smunu} we find
that the $t$- and $z$- integrals may be factorized (into $I_+$ and
$I_-$ respectively). To evaluate $I_+$ we express the $\cos(\Omega t)$
in its integrand in terms of exponentials, and perform a cylindrical
wave expansion via
\begin{align}
\label{cylindricalwaveexpansion}
e^{i\rho\cos\theta} =
\sum_{\ell = -\infty}^{+\infty} i^\ell J_\ell(\rho)
e^{i\ell\theta}.
\end{align}
The resulting Bessel functions in $I_+$ can be combined using
the recursion relation
\begin{align}
\label{besselrecursionnubyz}
\frac{\nu}{z} J_\nu(z) &= \frac{1}{2}\left( J_{\nu-1}(z) +
                               J_{\nu+1}(z) \right).
\end{align}
The integral for $I_-$ simply gives a $\delta$-function.
Together,
\begin{align}
\label{Iplusminus_infinitestring}
I_+^\alpha &= 2\pi \sum_{\ell = -\infty}^\infty
\delta\left( p_0 - \ell\Omega \right) (-1)^\ell
J_\ell\left( p_x \frac{v_0}{\Omega} \right)
\left( \delta^\alpha_0 - \delta^\alpha_x \frac{p_0}{p_x} \right)
\nonumber \\
I_-^\beta &= \delta^\beta_3 2\pi \delta(p_z)
\end{align}
and therefore
\begin{align*}
&{\bf I}_+ \times {\bf I}_- = {\bf \hat{y}} (2\pi)^2 \\
& \hskip 1 cm \times \sum_{\ell = -\infty}^\infty
\delta(p_z)
       \delta\left( p_0 - \ell\Omega \right) (-1)^\ell
J_\ell\left( p_x \frac{v_0}{\Omega} \right) \frac{p_0}{p_x} ~
\end{align*}
where we have used
\begin{align}
\label{besselntominusn}
J_{-n}(z) &= (-1)^n J_n(z), \quad n \in \{0,\pm 1, \pm 2,\dots\}
\end{align}

Making use of the formal identity
\begin{align*}
(\delta(p_z))^2 \to \delta(p_z) \frac{L}{2\pi}.
\end{align*}
where $L$ is the total length of the solenoid we find that
the fermion-antifermion pair production rate \eqref{pairproductionrate}
per unit length of the solenoid in sinusoidal motion is
\begin{equation}
{\dot{N'}} = \sum_{\ell=1}^{\infty} \int d^3 k \int d^3 k'  \sum_{s, s' = \pm} \frac{d^7 N (k s, k' s')}{d t d^3 k d^3 k'}
\label{eq:stringrate}
\end{equation}
where the differential rate
\begin{equation}
\frac{d^7 N (k s, k' s')}{d t d^3 k d^3 k'} = \frac{\epsilon^2}{(2 \pi)^2} \frac{{\cal N}( k s, k' s' )}{4 k_0 k_0' p_x^2}
J_{\ell}^2 \left( p_x \frac{v_0}{\Omega} \right) \delta (p_z) \delta (p^0 - \ell \Omega)
\label{eq:differentialrate}
\end{equation}
and $s$ and $s'$ represent the helicity of the particle and anti-particle respectively.
The quantity ${\cal N}$ is given by
\begin{eqnarray}
{\cal N} (k s, k's') & = & ( k_0 k_0' + m^2 - s s' | {\mathbf k} | | {\mathbf k'} | ) \nonumber \\
& \times & \left[ 1 + \frac{s s'}{| {\mathbf k} | | {\mathbf k'} |} ( k_z k_z' + k_x k_x' - k_y k_y' ) \right].
\nonumber \\
\label{eq:calnpol}
\end{eqnarray}
Note that in the limit that the particle and anti-particle are
ultra-relativistic ${\cal N} (k s, k' s')$ is negligible for
$s = s'$; thus pairs are predominantly produced with opposite
helicity in this limit. The pair production rate summed over
final state polarization is given by the simpler expression
\begin{eqnarray}
{\cal N} (k, k') & = & \sum_{s, s' = \pm} {\cal N} (k s, k' s')
\label{infinitestraightstring_pairproductionrate_numerator} \\
&& \hskip -.5 cm
=4( m^2 + k_0 k_0' - k_z k_z' - k_x k_x' + k_y k_y') \nonumber \\
&&\hskip -0.5 cm
\to 2\left( \left( \ell\Omega \right)^2 - \left( k_x +
k'_x \right)^2 - \left( k_y - k'_y \right)^2 \right) \nonumber
\end{eqnarray}
The second line in
\eqref{infinitestraightstring_pairproductionrate_numerator}
is the direct consequence of the term in curly brackets in
\eqref{pairproductionrate}. The third line, which is
$k_z$-independent,
has been obtained from the first by imposing $k_z = -k'_z$ and
re-writing $k_0 k'_0 = (1/2)(\ell\Omega)^2 - (1/2)k_0^2 -
(1/2)k_0'^2$,
using the constraints implied by the $\delta$-functions.

Note that when we square the amplitude we obtain a double sum over
harmonics. This collapses to a single sum because
$\delta(p_0-\ell\Omega) \delta(p_0-\ell'\Omega)$ is zero unless
$\ell = \ell'$, since the requirement that both the $\delta$-function
arguments be null cannot otherwise be satisfied. Since the sum of the
positive energies $p_0 = k_0 + k'_0$ cannot be zero or negative, we
have also removed all the $\ell \leq 0$ terms in the summation in
Eq.~(\ref{eq:stringrate}).

{\it Non-relativistic limit:} In the non-relativistic limit, $v_0 \ll 1$,
we may utilize
\begin{align}
\label{besselsmallz}
J_\nu(z) &\approx
\frac{1}{\Gamma(\nu+1)} \left(\frac{z}{2}\right)^\nu
\left( 1 + \mathcal{O}(z^2) \right), \qquad |z| \ll 1
\end{align}
to see that the contribution to the pair production rate at each harmonic
begins at $\mathcal{O}(v_0^{2\ell})$ plus corrections of
$\mathcal{O}(v_0^{2\ell+2})$. (For $\ell \gg 1$, the $J_\ell(z)$ becomes
exponentially suppressed; see equation \eqref{bessellargeindex}.)
Therefore, provided that $\Omega > 2m$, the first harmonic is the dominant
production channel in the non-relativistic limit.

We first, however, begin by performing a consistency check of our
calculations based on the $\ell = 1$ term. Using the pair production
rate in the form \eqref{pairproductionrate}, and keeping only the
leading order term in the series expansion of $J_1$ and eliminating
the $k_x k_x'$ and $k_y k'_y$ term by integrating over the appropriate
angular coordinates,
\begin{align}
\label{infinitestraightstring_pairproductionrate_NR}
\text{Eq }\eqref{eq:stringrate}
&\to \int_0^\infty dkk \int_0^\infty dk'k'
\int_{-\infty}^\infty dk_z \int_{-\infty}^\infty dk'_z \nonumber \\
&\qquad \times \delta(k_z+k'_z) \delta( k_0 + k'_0 - \Omega )
\frac{\epsilon^2}{k_0 k'_0} \frac{v_0^2}{4\Omega^2} \nonumber \\
&\qquad \times \left( m^2 + k_0 k'_0 + k_z^2 \right)
\end{align}
This will match \eqref{movingframespairproductionrate_smallABphase}
from the small flux limit of the moving frames perturbation theory
calculation below.

\begin{figure}
\scalebox{0.5}{\includegraphics{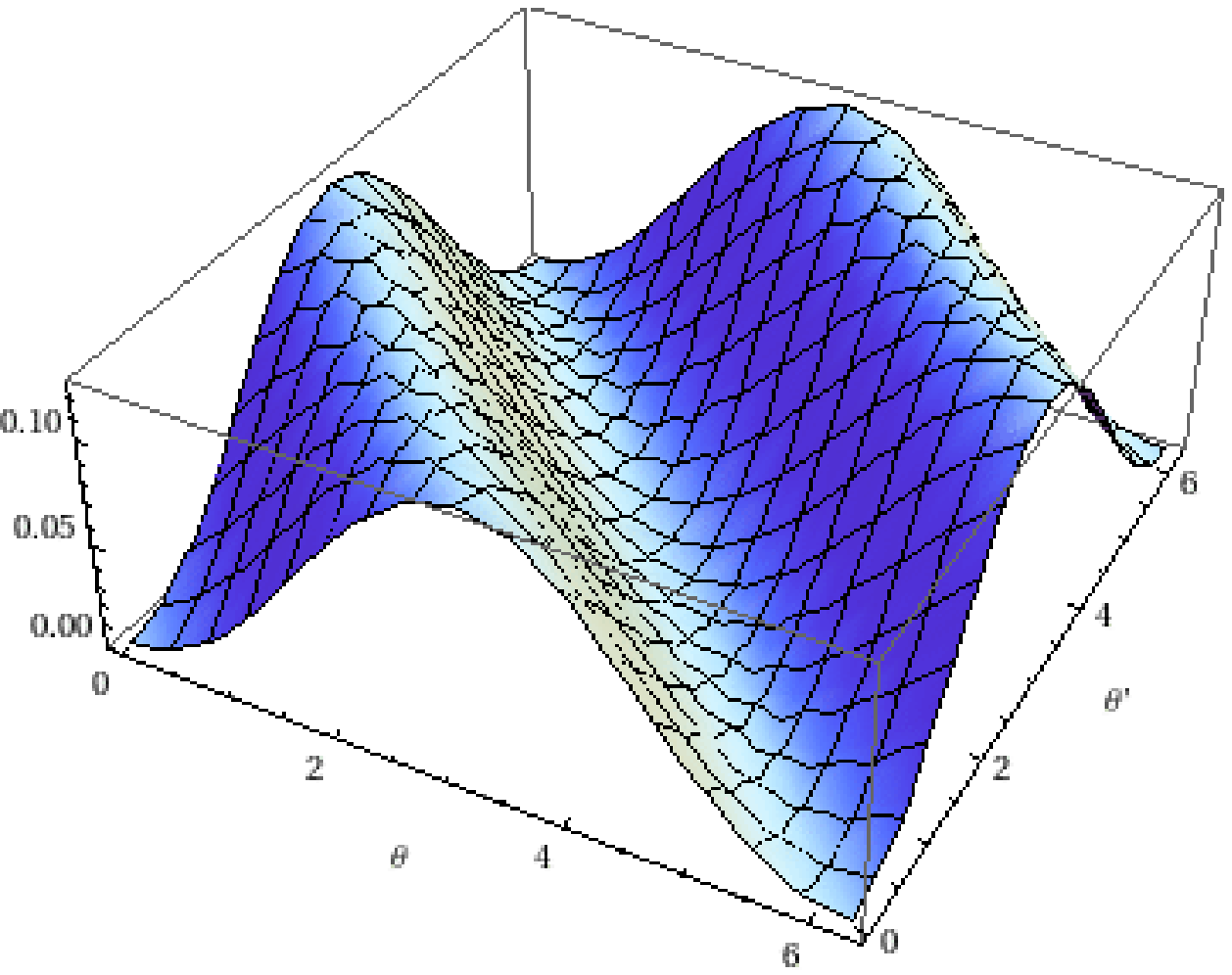}}
\scalebox{0.5}{\includegraphics{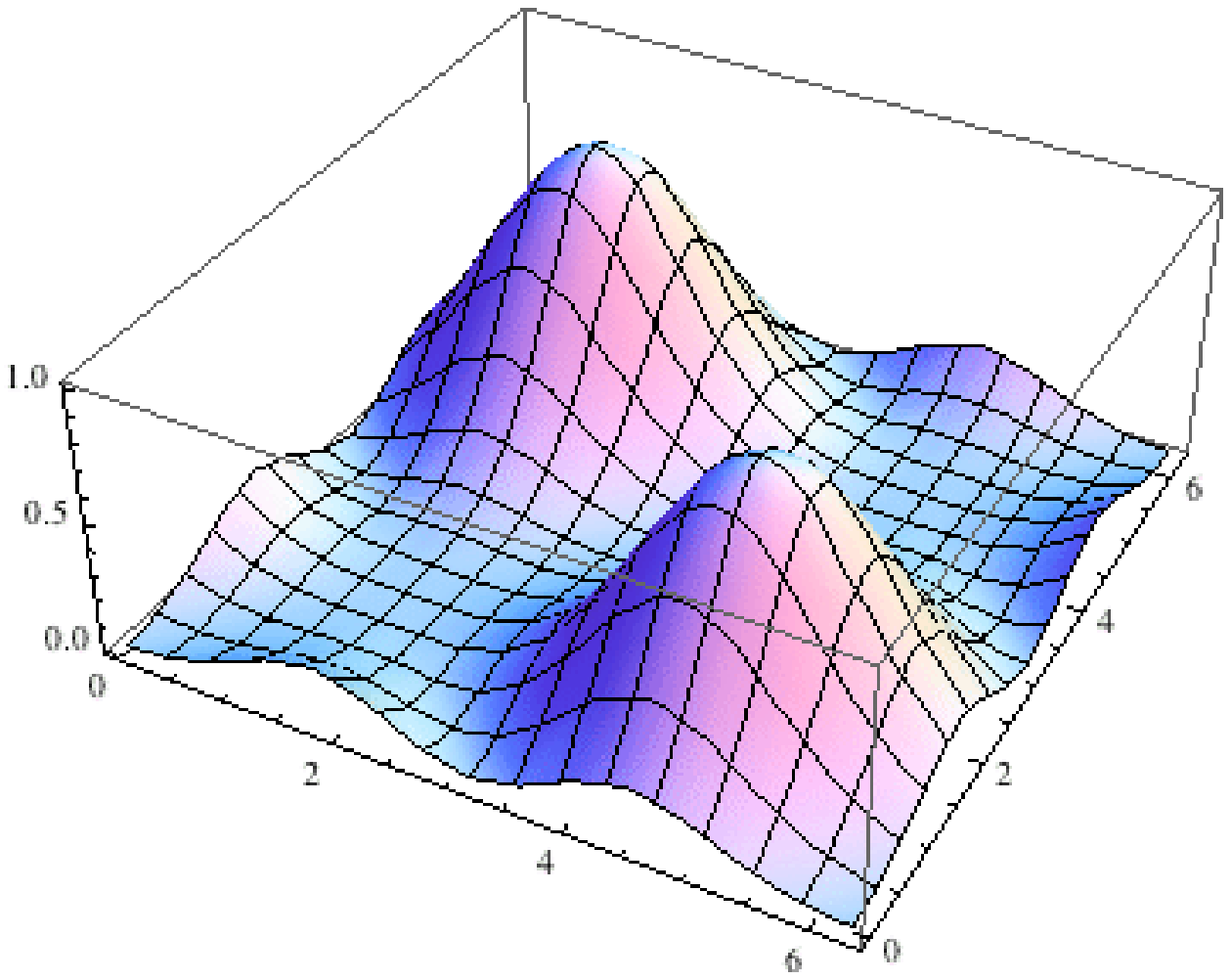}}
\caption{The radiated power as a function of
$\theta$ and $\theta '$. The top figure is for fermion
radiation $\ell=1$ with $|{\bm k}|=\ell \Omega/2 = |{\bm k}'|$,
and the lower figure is for bosonic radiation with
the same parameters. There is maximum radiation along the line
$\theta +\theta ' =\pi$ in the fermionic case and the
radiation is (approximately) circularly symmetric. The
bosonic emission is dipolar and along the $\theta+\theta'=2\pi$
line (plot taken from Ref.~\cite{AB_Bosons}). }
\label{angdistn}
\end{figure}

Now, the differential pair production rate is the integrand
in \eqref{eq:stringrate}. Upon integrating
$k_z$ (or $k'_z$), the third line in
\eqref{infinitestraightstring_pairproductionrate_numerator}
indicates that the only $k_z = -k'_z$ dependence in the integrand occurs in
the $k_0 k'_0$. Enforcing the constraint $k_0 + k'_0 = \Omega$
by introducing a Lagrange multiplier we see that the integrand,
and hence the emission rate, is maximum at $k_z = 0$. (This is
true of the full relativistic emission.)
Applying \eqref{besselsmallz} to $\ell = 1$ and going to cylindrical
coordinates, $(k_x, k_y) = k_\perp (\cos\theta, \sin\theta)$ and
$(k'_x, k'_y) = k'_\perp (\cos\theta', \sin\theta')$, together
with the second line of
\eqref{infinitestraightstring_pairproductionrate_numerator} then yields
\begin{align*}
&\frac{d^5 \dot{N}}{dk_\perp dk'_\perp d\theta d\theta' dk_z}
           (\ell = 1, k_z = k'_z = 0) \\
&= \left( \frac{v_0 \epsilon}{4\pi\Omega} \right)^2
  \frac{k_\perp k'_\perp}{k_0 k'_0}
            \delta\left( k_0 + k'_0 - \Omega \right) \\
&\qquad \times \left( m^2 + k_0 k'_0 -
         k_\perp k'_\perp \cos(\theta + \theta') \right) \\
\end{align*}
with
\begin{align*}
k_0 = \sqrt{{\bf k}^2_\perp+m^2}, \qquad k'_0 = \sqrt{{\bf k}'^2_\perp+m^2}
\end{align*}
The differential rate is largest in the $xy$-plane and when the
outgoing particles' azimuthal angles are supplementary,
$\theta + \theta' = \pi$.
In contrast, boson emission is maximum along\footnote{This
corrects \cite{AB_Bosons} where it was mistakenly stated
that the maximum emission is along $\theta+\theta'=\pi$.}
$\theta+\theta'=2\pi$ and dipolar (see Fig.~\ref{angdistn}).

We may proceed to employ the second line of
\eqref{infinitestraightstring_pairproductionrate_numerator} to obtain
the total rate of production of pairs with energy $\Omega$ per unit
length of the infinite flux tube
{\allowdisplaybreaks
\begin{align*}
&{\dot N}'(\ell =1)
= \frac{\epsilon^2}{2(2\pi)^2} \left(\frac{v_0}{2\Omega} \right)^{2} \\
&\times \int_{m}^{\Omega-m} dk_0
        \int_{-\sqrt{k_0^2-m^2}}^{\sqrt{k_0^2-m^2}} dk_z
        \int_0^{2\pi} d\theta \int_0^{2\pi} d\theta' \\
&\times [ \Omega^2 - (k_x+k'_x)^2 - (k_y - k'_y)^2 ],
\end{align*}}
where
\begin{eqnarray}
(k_x,k_y) &=& \sqrt{k_0^2 - m^2 - k_z^2} ~
                       (\cos\theta,\sin\theta) \nonumber\\
(k'_x,k'_y) &=& \sqrt{(\Omega-k_0)^2 - m^2 - k_z^2} ~
                       (\cos\theta',\sin\theta') \nonumber
\end{eqnarray}
In terms of the mass-to-energy ratio $\tau \equiv 2m/\Omega$, the
result is
\begin{align}
&{\dot N}'(\ell=1) =
\frac{v_0^2 \epsilon^2 \Omega^2}{384}
\bigg( 2 \sqrt{1-\tau} [16 - \tau \{ \tau (3 \tau +2)+8 \}] \nonumber\\
  &\qquad - 6\tau^4 \ln\left[ \sqrt{\frac{1}{\tau}-1} +
     \sqrt{\frac{1}{\tau}} \right] \bigg) + \mathcal{O}(v_0^4)
\label{pprodrateleq1}
\end{align}
This pair production rate begins at $v_0^2 \epsilon^2 \Omega^2/12$
for $\tau = 0$ and, for increasing $\tau$, decreases monotonically
to zero at the threshold $\tau = 1$.
As the fermion-antifermion pair gets heavier, they become harder
to produce.

\begin{figure}
\scalebox{0.7}{\includegraphics{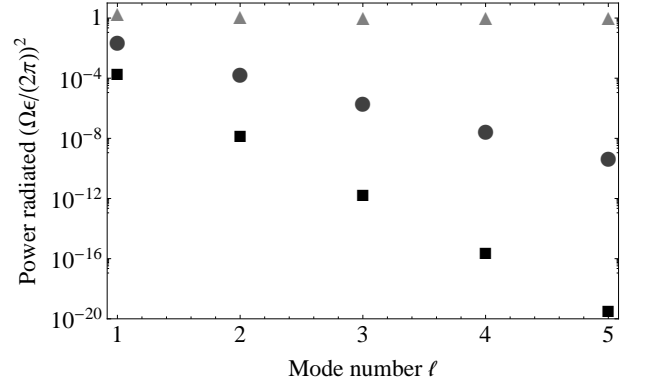}}
\caption{The radiated power as a function of harmonic for
$v_0=0.001$ (squares), $0.1$ (circles) and $1$ (triangles).
}
\label{totalpower}
\end{figure}

At very high harmonics ($\ell \gg 1$), we may re-scale all
momenta by $\ell\Omega/v_0$ and denote the new variables by
overbars, {\it e.g.} $\bar{k} \equiv k v_0/(\ell\Omega)$.
Then we invoke
\cite{MorseFeshbachWatson}
\begin{align}
\label{bessellargeindex}
J_\nu\left( \frac{\nu}{\cosh\alpha} \right) &\sim
\frac{\exp\left( - \nu (\alpha- \tanh\alpha) \right)}
     {\sqrt{2\pi\nu \tanh\alpha}}, \ \ \alpha > 0, \ \ \nu \gg 1,
\end{align}
to say that, within the integration limits
$0 < \bar{p}_x \lesssim v_0 \ll 1$,
\begin{align}
\label{infinitestraightstring_besselJ_exp}
\left( \frac{J_\ell(\ell \bar{p}_x)}{\bar{p}_x} \right)^2 \sim
\frac{\exp[ - 2 \ell (\alpha_0 - \tanh\alpha_0) +
    2 \ln( \cosh\alpha_0 )]}{\sqrt{2\pi\ell\tanh\alpha_0}},
\end{align}
where $\alpha_0$ is the positive solution to the equation
$\cosh\alpha_0 = 1/\bar{p}_x$.

The exponent on the right hand side of the asymptotic expression
\eqref{infinitestraightstring_besselJ_exp} is a monotonically
increasing function of $\bar{p}_x$, and takes on large negative
values for small $\bar{p}_x$ ({\it i.e.} large $\alpha_0$), while
$\tanh\alpha_0 \sim 1$. Hence the pair production rate
\eqref{eq:stringrate} is exponentially
small for $\ell \gg 1$, in the non-relativistic limit.
In Fig.~\ref{totalpower} we show the power emitted per harmonic
for a few values of $v_0$.

{\it High frequency limit:} When $\Omega \gg 2m$, such that the
fermion-antifermion pairs are produced with very large momentum,
to a good approximation, they may be treated as effectively massless.
If we re-scale all momenta by the energy, $k \equiv \ell \Omega \bar{k}$
etc., and integrate over $\bar{k}'_\perp$, the emission rate in the
$xy$-plane is
\begin{align*}
&\frac{d^4 \dot{N}}{d\bar{k}_\perp d\theta d\theta' dk_z}
(k_z = k'_z = 0, m = 0) \\
&= \left( \frac{\ell\Omega\epsilon}{2\pi}
\frac{J_\ell( \ell v_0 \bar{p}_x)}{\bar{p}_x} \right)^2
\bar{k}_\perp (1-\bar{k}_\perp) \left( 1 - \cos(\theta+\theta') \right),
\end{align*}
with
\begin{align*}
&\bar{p}_x = \bar{k}_\perp \cos\theta + (1-\bar{k}_\perp) \cos\theta'
\end{align*}
The rescaled momentum, $\bar{k}_\perp$, lies in (0,1) and
so $\bar{p}_x$ lies in the interval $(\cos\theta,\cos\theta')$
for $\cos\theta < \cos\theta'$, and $(\cos\theta',\cos\theta)$
for $\cos\theta > \cos\theta'$. Therefore the absolute value of
the argument of $J_\ell$ is always less than its order,
$|\ell v_0 \bar{p}_x| \leq \ell v_0 < \ell$. For $\ell = 1$, as a first
approximation, we may recall \eqref{besselsmallz}, and observe that
$(J_\ell(\ell v_0 \bar{p}_x)/\bar{p}_x)^2$ stays roughly constant
within the range of interest at hand. Hence, the maximum emission
occurs for $\bar{k}_\perp \approx 1/2$ and
$\theta + \theta' =\pi$. For $\ell \geq 2$, we may employ
\eqref{besselrecursionnubyz} followed by the fact that both the first turning point and zero of $J_\ell(z)$ occurs only for
$z \geq \ell$, to argue that $(J_\ell(\ell v_0 \bar{p}_x)/\bar{p}_x)^2$
is a monotonically increasing function of $\bar{p}_x$ -- this means
the $\bar{k}_\perp$ at which there is maximum emission now shifts
away from $1/2$, with the direction depending on whether
$\cos\theta$ is greater or less than $\cos\theta'$.

At higher harmonics, $\ell \gg 1$, $J_\ell$ becomes exponentially
suppressed, according to \eqref{bessellargeindex} and
\eqref{infinitestraightstring_besselJ_exp}, in most of the
$(\theta,\theta')$ plane. If the motion is relativistic
$(v_0 \sim 1)$, however, the asymptotic formula \cite{MorseFeshbachWatson}
\begin{eqnarray}
\label{bessellargeindex_nueqz}
J_\ell(z) &\sim& \frac{1}{3\pi} \biggl(
  \frac{\sin(\pi/3) \Gamma(1/3)}{(z/6)^{1/3}} \\
  && +\frac{\sin(2\pi/3) \Gamma(2/3)}{(z/6)^{2/3}} (z-\ell)
        + \dots \biggr), \qquad \ell \gg 1, \nonumber
\end{eqnarray}
valid for $z \sim \ell$, says that
$(J_\ell(\ell v_0 \bar{p}_x)/\bar{p}_x)^2$ will transition to an
inverse power law in $\ell$ and hence peak in the region where
$|\bar{p}_x|$ lies closest to unity, namely, where
$\cos\theta = \cos\theta' = \pm 1$. (Note that
$(J_\ell(-z)/(-z))^2 = (J_\ell(z)/z)^2$.) Therefore, ignoring
the $(1-\cos(\theta+\theta'))$ factor for now, the peak
occurs at $(\theta ,\theta')=(0,0)$ or $(\pi,\pi)$.
The shape of the peak
is determined by the constant $\bar{p}_x$ contour lines on the
$(\theta,\theta')$ plane near $(0,0)$ and $(\pi,\pi)$. They are
ellipses with $\bar{k}_\perp$-dependent eccentricity, because
\begin{equation}
\bar{p}_x(0,0) = 1 - \frac{1}{2}(\bar{k}_\perp \theta^2 +
(1-\bar{k}_\perp) \theta'^2) + \dots
\nonumber
\end{equation}
\begin{equation}
\bar{p}_x(\pi,\pi) = -1 + \frac{1}{2}(\bar{k}_\perp (\theta-\pi)^2 +
(1-\bar{k}_\perp) (\theta'-\pi)^2) + \dots
\nonumber
\end{equation}
Now we include the effect of the $(1-\cos(\theta+\theta'))$ factor
which vanishes along the lines $\theta + \theta' = 2\pi n$, where
$n \in \mathbb{Z}$. The peaks at $(0,0)$ and $(\pi,\pi)$ lie
precisely on these lines. So the multiplicative factor
$(1-\cos(\theta+\theta'))$ modulates the peak and splits it
into two, one on either side of the $\theta + \theta' = 2\pi n$
lines.

To summarize, at large oscillation frequencies and relativistic
solenoid speeds, we expect the high energy spectrum of the
fermion-antifermion pairs to be produced predominantly in
the $xy$-plane, in narrow pairs of nearly collinear beams,
in the $\pm \widehat{x}$ directions. This beaming effect is due to
the presence of the Bessel functions $J_\ell(\ell v_0 \bar{p}_x)$,
common to both the fermion and boson amplitudes, and hence
will occur in the boson case too \cite{AB_Bosons}.

\section{Cosmic String Loops}
\label{cosmicstringloops}

We now turn to fermion pair production from cosmic string loops
for which the dynamics is given by the Nambu-Goto action
\begin{equation}
\label{nambugotoaction}
S_\text{NG} \equiv -\mu \int d\tau \int d\sigma \sqrt{-\gamma}
\end{equation}
where the world-sheet metric is
\begin{equation}
\gamma_{ab} = \eta_{\mu\nu} \partial_a X^\mu \partial_b X^\nu \ , \ \
\gamma \equiv \det(\gamma_{ab})
\end{equation}
and $\zeta^a =(\tau,\sigma)$ are world-sheet coordinates.

Referring to the Nambu-Goto dynamics for relativistic strings in
\eqref{nambugotoaction}, we will choose $\sigma$ such that constant
$t$- and $\sigma$-lines on the string world sheet are orthogonal,
and the energy per unit (proper) length of the string is constant,
\begin{align}
\label{worldsheetgaugeconditions}
X' \cdot \dot{X} = 0, \qquad X'^2 + \dot{X}^2 = 0.
\end{align}
These choice of coordinate conditions \eqref{worldsheetgaugeconditions}
put the world sheet metric $\gamma_{ab}$ into a conformally flat form,
and lead us from the Nambu-Goto action \eqref{nambugotoaction} to the
wave equation for $X^\mu$
\begin{align}
\label{nambugoto_waveequation}
\left( \partial_t^2 - \partial_\sigma^2 \right) X^\mu(t,\sigma) = 0
\end{align}
The $X^\mu$ solution may be written as an average of left- and
right-movers, $L^\mu = L^\mu(\sigma+t)$ and $R^\mu = R^\mu(\sigma-t)$
respectively,
\begin{align*}
X^\mu = \frac{1}{2} L^\mu(\sigma_+) + \frac{1}{2} R^\mu(\sigma_-)
\end{align*}
where we also have introduced world sheet light cone coordinates
\begin{align*}
\sigma_\pm \equiv \sigma \pm t,
\end{align*}
and
\begin{align*}
L^0 = \sigma_+, \quad R^0 = -\sigma_-.
\end{align*}
The world sheet coordinate conditions \eqref{worldsheetgaugeconditions}
translate into constraints on the spatial components of $X^\mu$,
\begin{align*}
{\bf L}'^2 = {\bf R}'^2 = 1
\end{align*}

We will focus our attention on two specific loop configurations.
The kinky loop has tangent vectors to the string that are discontinuous
at isolated points; whereas the cuspy loop has isolated points that
reach the speed of light periodically. These configurations satisfy
the equations of motion \eqref{nambugoto_waveequation} that follow
from \eqref{nambugotoaction}.

If the coordinate length of a given cosmic string loop is $L$,
the $\sigma$-integration limits for its corresponding
$S^{\mu\nu}$, (see Eq.~\eqref{Smunu}) runs from $0$ to $L$.
Since we have closed loops, ${\bf X}$ and derivatives
$\partial_t X^\mu$ and $\partial_\sigma X^\mu$ are periodic
in $\sigma$ as well as $t$. To facilitate the computation, we
utilize a formal device introduced in \cite{AB_Bosons} that
allows us to factorize $S^{\mu\nu}$ into a product of two
one-dimensional integrals, one for each world sheet light cone
coordinate. This is based on the observation that, for a periodic
function $f$ with period $L$, the integral over one period of $f$
is equivalent to the integral over the real line, divided by the
(infinite) number of times the former has been over-counted
\begin{equation}
\label{overcountingtrick}
\int_0^L d\sigma f(\sigma) =
\frac{1}{\delta_\mathbb{Z}(0)}
           \int_{-\infty}^{+\infty} d\sigma f(\sigma),
\end{equation}
where
\begin{equation*}
\delta_\mathbb{Z}(0) \equiv
\frac{\int_{-\infty}^{+\infty} d\sigma e^{i2\pi\ell\sigma /L}}
{\int_0^L d\sigma e^{i2\pi\ell\sigma /L}},
\qquad \ell \in \mathbb{Z}
\end{equation*}
We may now extend the $\sigma$ integral in $S^{\mu\nu}$ to the
entire real line using \eqref{overcountingtrick}, before changing
variables from $(t,\sigma)$ to $(\sigma_+,\sigma_-)$.
$S^{\mu\nu}$ then factorizes into
\begin{align}
\label{IplusIminus}
S^{\mu\nu} &= \frac{1}{2} I_+^{[\mu} I_-^{\nu]} \\
I_+^\alpha &\equiv \frac{1}{2}
\int_{-\infty}^{+\infty} d\sigma_+
\partial_+ L^\alpha e^{ip \cdot L/2} \nonumber \\
I_-^\alpha &\equiv \frac{1}{2\delta_\mathbb{Z}(0)}
\int_{-\infty}^{+\infty} d\sigma_-
\partial_- R^\alpha e^{ip \cdot R/2}, \nonumber
\end{align}
where the derivatives are with respect to $\sigma_\pm$\footnote{If
we had not applied \eqref{overcountingtrick}, upon converting
to light cone coordinates, the inner $\sigma_\pm$-integration
would have limits that depend on the outer integration variable,
and $S^{\mu\nu}$ would not factorize.}.
Periodicity in $\sigma$ and $t$ implies that the integrands in
\eqref{IplusIminus} are periodic in $\sigma_+$ and $\sigma_-$ and
we can replace them in Eq.~(\ref{Smunu}) with their discrete Fourier
series expansions,
\begin{align}
\label{I_pm_fourierseries}
I_+^\alpha &= 2\pi \sum_{\ell = -\infty}^{\infty}
     \delta\left( p_0 + \frac{4\pi}{L} \ell \right) \nonumber \\
&\qquad \times \int_0^L \frac{d\sigma_+}{L}
\partial_+ L^\alpha e^{-i\ell 2\pi\sigma_+/L}
e^{-i{\bf p} \cdot {\bf L}/2} \nonumber \\
\end{align}
\begin{align}
I_-^\alpha &= \frac{2\pi}{\delta_\mathbb{Z}(0)}
       \sum_{\ell = -\infty}^{\infty}
        \delta\left( p_0 - \frac{4\pi}{L} \ell \right) \nonumber \\
&\qquad \times \int_0^L \frac{d\sigma_-}{L}
\partial_- R^\alpha e^{-i2\pi\ell\sigma_-/L}
e^{-i {\bf p} \cdot {\bf R}/2} \nonumber
\end{align}

The ${\bf I}_+ \times {\bf I}_-$ that follows from
\eqref{I_pm_fourierseries} will be an infinite sum involving
$(\delta(p_0 - \ell 4\pi/L))^2$. We write
\begin{align}
\label{deltafunctionsquared_Z}
(\delta(p_0 - \ell 4\pi/L))^2 =
\delta(p_0 - \ell 4\pi/L) \delta_\mathbb{Z}(0) \frac{L}{4\pi}
\end{align}
and the $\delta_\mathbb{Z}(0)$ in
\eqref{deltafunctionsquared_Z} will cancel that in $I_-$
\eqref{I_pm_fourierseries}.

\subsection{Kinky Loops}
\label{kinkyloops}

The ``degenerate" kinky loop solution we will consider is
\begin{align}
\label{LR_kinkyloop}
{\bf L}(\sigma_+) &= \left\{ \begin{array}{ll} \sigma_+
{\bf A} & 0 \leq \sigma_+ \leq \frac{L}{2} \\
(L-\sigma_+) {\bf A} & \frac{L}{2} \leq \sigma_+ \leq L \\
\end{array} \right. \nonumber \\
{\bf R}(\sigma_-) &= \left\{ \begin{array}{ll} \sigma_-
{\bf B} & 0 \leq \sigma_- \leq \frac{L}{2} \\
(L-\sigma_-) {\bf B} & \frac{L}{2} \leq \sigma_- \leq L \\
\end{array} \right.
\end{align}
where ${\bf A}$ and ${\bf B}$ are unit vectors.
This loop is "degenerate" because it consists of four
straight segments and is "kinky" because of its four corners.
The four straight segments propagate with constant
speed but shrink and expand due to the motion of the
kinks.

Now, denoting
\begin{align*}
p_\text{A} \equiv p_i {\bf A}^i, \qquad
p_\text{B} \equiv p_i {\bf B}^i,
\end{align*}
and putting the kinky loop trajectory \eqref{LR_kinkyloop}
into \eqref{I_pm_fourierseries} then leads to
\begin{align}
\label{Iplusminus_kinkyloop}
I_+^\alpha &= \sum_{\ell = -\infty}^{\infty}
\left( p_\text{A}, -p_0 {\bf A} \right)
\delta\left( p_0 - \frac{4\pi}{L} \ell \right) \nonumber \\
&\qquad \times \frac{16\pi e^{i(p_\text{A}L/8-\pi\ell /2 )}}
     {L(p_\text{A}^2 - p_0^2)}
    \sin\left( \frac{p_\text{A}}{8}L-\frac{\pi}{2}\ell \right) \nonumber \\
I_-^\alpha &= \frac{1}{\delta_\mathbb{Z}(0)}
 \sum_{\ell = -\infty}^{\infty} \left( -p_\text{B}, p_0 {\bf B} \right)
\delta\left( p_0 - \frac{4\pi}{L} \ell \right) \nonumber \\
&\qquad \times \frac{16\pi e^{i(p_\text{B}L/8+\pi \ell/2)}}
                     {L(p_\text{B}^2 - p_0^2)}
\sin\left( \frac{p_\text{B}}{8}L+\frac{\pi}{2}\ell \right)
\end{align}
and
\begin{align*}
&{\bf I}_+ \times {\bf I}_-
      = -16 \sum_{\ell = -\infty}^{\infty} p_0^2
\frac{4\pi}{L} \frac{e^{i(p_\text{A}+p_\text{B})L/8}}
    {(p_\text{A}^2 - p_0^2)(p_\text{B}^2 - p_0^2)}\nonumber \\
&\qquad \times \sin\left( \frac{p_\text{A}}{8}L-\frac{\pi}{2}\ell \right)
  \sin\left( \frac{p_\text{B}}{8}L+\frac{\pi}{2}\ell \right) \\
&\qquad \times \delta\left( p_0 - \frac{4\pi}{L} \ell \right)
{\bf A} \times {\bf B}
\end{align*}
The power radiated from the kinky loop due to fermion pairs emitted
with energy $\ell\Omega$, for a fixed $\ell$, is the corresponding
pair production rate multiplied by the energy $4\pi\ell/L$,
\begin{align}
\label{powerradiated_kinkyloop_ell}
&\dot{E}^{(\text{K})}_\ell = 2\pi \epsilon^2
    \left( \frac{64\pi}{L} \right)^2 \left( \frac{4\pi\ell}{L} \right)^3
\int \frac{d^3 k}{(2\pi)^3} \frac{d^3 k'}{(2\pi)^3} \frac{1}{k_0 k'_0}
\nonumber\\
&\times \frac{\sin^2\left( {p_\text{A}L}/8-{\pi \ell}/{2} \right)
              \sin^2\left( {p_\text{B}L}/8+{\pi \ell}/{2} \right)}
 {(p_\text{A}^2 - p_0^2)^2(p_\text{B}^2 - p_0^2)^2} \nonumber\\
& \times [ ({\bf A} \times {\bf B} )^2
     \left( m^2 + k \cdot k' \right) +
2  {\bf A} \times {\bf B} \cdot {\bf k} ~
  {\bf A} \times {\bf B}  \cdot {\bf k}' ]
\nonumber \\
  & \times \delta\left( p_0 - \frac{4\pi}{L} \ell \right)
\end{align}
The total power radiated is then
\begin{align*}
\dot{E}^{(\text{K})}(N) = \sum_{\ell = 1}^N \dot{E}^{(\text{K})}_\ell
\end{align*}
We have truncated the summation at some large integer $N$ because it
will turn out, just like in the bosonic case \cite{AB_Bosons}, that the
total power obtained from summing to $\ell = \infty$ will diverge.
The cut-off is related to the rounding off of the kink and may
be estimated as the ratio of the length of the cosmic string loop,
$L$, to its width $w$. Taking $L \sim 1$ Mpc and
$w \sim 1 \text{ TeV}^{-1}$, we obtain $N \sim 10^{41}$. For the
electron -- the lightest electrically charged fermion -- note that
the relevant range of mode number is
$10^{35} \lesssim \ell_{e^+e^-} \lesssim 10^{41}$ for such an $L$ and
$w$, where the lower limit is determined by the product of the electron
mass and $L$. In the high energy range of the spectrum, which gives
the dominant contribution to the power emitted, the fermions can be
treated as effectively massless.

\begin{figure}
\begin{center}
\includegraphics[width=3.4in]{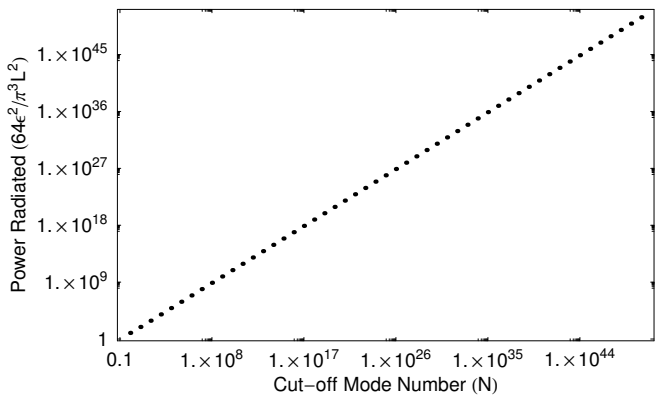} \\
\end{center}
\caption{Plot of power radiated in the form of massless
fermion-antifermion pairs in units of $64\epsilon^2/(\pi^3L^2)$,
from the degenerate kinky loop with ${\bf A}$ and ${\bf B}$
perpendicular, as a function of the cut-off mode number $N$.}
\label{KinkyIntegralPlot}
\end{figure}

When $m=0$, by re-scaling the momenta in the integral
\eqref{powerradiated_kinkyloop_ell}
via $(k,k') \equiv (4\pi\ell/L) (\bar{k},\bar{k}')$,
the sole dependence on the summation index $\ell$ occurs
in the trignometric functions, which in turn can be summed using the
formula
\begin{align*}
&\sum_{\ell = 1}^N \sin^2(\ell x) \sin^2(\ell y) =
\frac{M}{8} - \frac{\sin(M x)}{8 \sin x} - \frac{\sin(M y)}{8
\sin y} \\
&\quad + \frac{\sin(M(x-y))}{16 \sin(x-y)} +
\frac{\sin(M(x+y))}{16 \sin(x+y)}, \ \ M \equiv 2N+1
\end{align*}
The resulting integral was then evaluated numerically with
{\sf Mathematica}.

In Fig. \ref{KinkyIntegralPlot} we have plotted the total power
$\dot{E}(N)$ as a function of the cut-off $N$, for $m=0$ and
for the ``square'' loop which has ${\bf A}\cdot {\bf B}=0$,
from $N = 1$ up to $N = 10^{50}$. We see that the power radiated
per mode $\ell$ is indeed independent of $\ell$, in this massless
limit, and hence the total power emitted grows linearly
with $N \sim L/w$ and can be very large.

\subsection{Cuspy Loops}
\label{cuspyloops}

The cuspy loop solution we will consider is
\begin{align}
\label{LR_cuspyloop}
{\bf L}(\sigma_+) &=
\frac{L}{2\pi}\left( \sin\left( \frac{2\pi\sigma_+}{L} \right), 0,
-\cos\left( \frac{2\pi\sigma_+}{L} \right) \right) \nonumber \\
{\bf R}(\sigma_-) &=
\frac{L}{2\pi}\left( \sin\left( \frac{2\pi\sigma_-}{L} \right),
-\cos\left( \frac{2\pi\sigma_-}{L} \right), 0 \right)
\end{align}
The loop has cusps because there are points such that
$\partial_+ {\bf L} = - \partial_- {\bf R}$. These occur at
$2\pi\sigma_\pm/L = 0, \pi$ at which point the velocity of the
string is $\pm \hat{\bf x}$, {\it i.e.} the point on the string
reaches the speed of light.

Putting the cuspy loop trajectory \eqref{LR_cuspyloop} into
\eqref{I_pm_fourierseries}, combining the sines and cosines
occurring in the exponential into one trigonometric function
before performing a cylindrical wave expansion with
\eqref{cylindricalwaveexpansion} then gives us
\begin{align}
\label{Iplusminus_cuspyloop}
I_+^\alpha &=
2\pi \sum_{\ell = -\infty}^\infty \left( \varphi_{(+|-\ell)},
i \frac{4\pi}{L} \hat{\bf y} \times \vec{\partial}_p
\varphi_{(+|-\ell)} \right)
\delta\left( p_0 - \frac{4\pi}{L} \ell \right) \nonumber \\
I_-^\alpha &=
\frac{2\pi}{\delta_\mathbb{Z}[0]}
\sum_{\ell = -\infty}^\infty
\left( \varphi_{(-|\ell)}, i \frac{4\pi}{L} \hat{\bf z} \times
\vec{\partial}_p \varphi_{(-|\ell)} \right)
\delta\left( p_0 - \frac{4\pi}{L} \ell \right)
\end{align}
with
\begin{align*}
\varphi_{(+|\ell)}(p_x,p_z) &=
i^\ell J_\ell\left( \frac{L}{4\pi} \sqrt{p_x^2+p_z^2} \right)
\exp\left(i\ell \arctan\left( \frac{p_x}{p_z} \right) \right) \\
\varphi_{(-|\ell)}(p_x,p_y) &=
-i^\ell J_\ell\left( \frac{L}{4\pi} \sqrt{p_x^2+p_y^2} \right)
\exp\left(i \ell \arctan\left( \frac{p_x}{p_y} \right) \right) \\
\vec{\partial}_p &\equiv
\left( \frac{\partial}{\partial p_x}, \frac{\partial}{\partial p_y},
\frac{\partial}{\partial p_z} \right)
\end{align*}
as well as
\begin{align*}
{\bf I}_+ \times {\bf I}_-
&= - \frac{16\pi^3}{L} \sum_{\ell = -\infty}^\infty
\delta\left( p_0 - \frac{4\pi}{L} \ell \right) {\bf n}_\ell
\end{align*}
with
\begin{align*}
{\bf n}_\ell \equiv
\left( \hat{\bf y} \times \vec{\partial}_p \varphi_{(+|-\ell)} \right)
\times \left( \hat{\bf z} \times \vec{\partial}_p
\varphi_{(-|\ell)} \right)
\end{align*}
The derivatives on $J_\ell$ occuring in ${\bf I}_\pm$ and
${\bf n}_\ell$ may be carried out with the aid of
\begin{align}
\label{besselrecursiondz}
\partial_z J_\nu(z) &= \frac{1}{2}\left( J_{\nu-1}(z) - J_{\nu+1}(z) \right)
\end{align}
The power radiated from the cuspy loop due to fermion pairs emitted
with energy $\ell\Omega$, for a fixed $\ell$, is the corresponding
pair production rate multiplied by the energy $4\pi\ell/L$,
\begin{align}
\label{powerradiated_cuspyloop_ell}
\dot{E}^{(\text{C})}_\ell &=
\int d^3 k \int d^3 k' \frac{2}{\ell L} \frac{\epsilon^2}{k_0 k'_0}
\delta\left( p_0 - \frac{4\pi}{L} \ell \right) \nonumber\\
&\hskip -1 cm \times \bigg\{ \left\vert {\bf n}_\ell \right\vert^2
\left( m^2 + k \cdot k' \right) + \left( {\bf n}_\ell \cdot {\bf k}
\ {\bf n}^*_\ell \cdot {\bf k}' + \text{c.c.} \right) \bigg\}
\end{align}

\begin{figure}
\begin{center}
\includegraphics[width=3.4in]{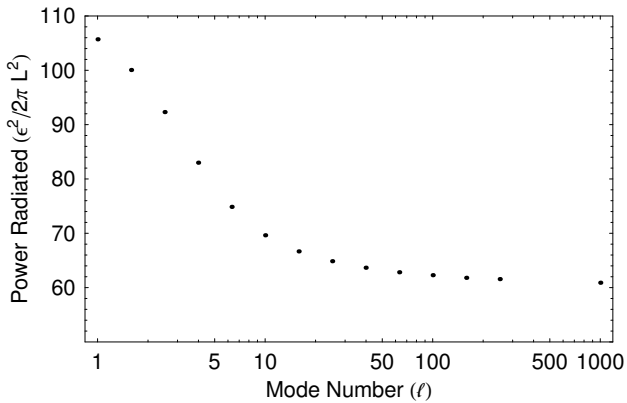} \\
\end{center}
\caption{Plot of power radiated in the form of massless fermion-antifermion
pairs in units of $\epsilon^2/(2\pi L^2)$, from the cuspy loop, as a
function of the mode number $l$.}
\label{CuspyyIntegralPlot}
\end{figure}

Using {\sf Mathematica}, we evaluated \eqref{powerradiated_cuspyloop_ell}
numerically for $\ell = 1$ through $\ell = 1000$ and $m=0$.
Fig.~\ref{CuspyyIntegralPlot} displays the resulting power radiated in
the form of massless fermions at each harmonic, with energy $4\pi \ell/L$.
This provides evidence that, at large mode numbers ($\ell \gg 1$), the
power radiated varies very slowly with $\ell$; though a thorough analysis
would have to employ more sophisticated numerical techniques (or
semi-analytic ones, using \eqref{bessellargeindex} and
\eqref{bessellargeindex_nueqz}) to evaluate
\eqref{powerradiated_cuspyloop_ell} for the astrophysically relevant
range of $10^{30} \lesssim \ell \lesssim 10^{50}$.

\section{Infinite, Straight Solenoid via Moving Frames Perturbation Theory}
\label{movingframesperturbationtheory}

In the second half of this paper, we provide an alternate calculation
of the fermion pair production rate from the infinite straight solenoid,
introduced in Sec.~\ref{infinitestraightstring}, aligned with the
$z$-axis, moving in sinusoidal motion along the $x$-axis. The spatial location of the solenoid as a function of time $t$ is given by
${\bf X}(t,z) = (\xi(t),0,z)$, with $\xi(t) = (v_0/\Omega)\sin(\Omega t)$.
This non-relativistic computation will only capture the first harmonic
of the infinite sum obtained in Sec.~\ref{infinitestraightstring} --
see \eqref{movingframestimeintegral} below -- but will make manifest
the periodic dependence of the pair production rate on the AB phase
$\epsilon = e\Phi/2\pi$ that is expected from such a topological
interaction. It will also serve as a consistency check on the results
in Sec.~\ref{infinitestraightstring}.

Let $\chi_{(s|+)}({\bf x}) e^{- iE_s t}$ denote the positive and
$\chi_{(s|-)}({\bf x}) e^{+ iE_s t}$ the negative energy solutions to
the Dirac equation in the presence of the gauge potential $A_\mu$ of
a static, infinite straight solenoid aligned along the $z$-axis centered
at $(x,y) = (0,0)$. (The subscripts $s$ and $s'$ in this section denote,
collectively, all the rest of labels that come with the solutions.)
We will expand the Dirac operator, $\psi$, in terms of ``shifted''
mode functions, $\chi_{(s|\pm)}(x-\xi,y,z) e^{\mp iE_s t}$. Then
we can show, as derived in Sec.~\ref{movfraint}, that the
fermion-antifermion pair production amplitude for a solenoid moving
along the $x$-axis is provided by the product of the integrals
\begin{align}
\label{movingframesintegral}
{}_{\rm out}\langle s,s' | 0 \rangle_{\rm in} &
\approx - \int_{-\infty}^{+\infty} dt' \dot{\xi}(t')
e^{i(E_{s'}+E_s)t'} \nonumber \\
&\times \int d^3 x' \chi_{(s|+)}^\dagger({\bf x}')
\partial_x \chi_{(s'|-)}({\bf x}'),
\end{align}
where the left-hand side of the equation is an inner product in
the Heisenberg picture.

If we specialize to the sinusoidal trajectory considered in
Sec.~\ref{infinitestraightstring}, the time integral in
\eqref{movingframesintegral} can be evaluated immediately to
yield conservation of energy
\begin{align}
\label{movingframestimeintegral}
\int_{-\infty}^{+\infty} dt' \dot{\xi}(t') e^{i(E_{s'}+E_s)t'}
&= -i\pi v_0 \delta(E_s + E_{s'} - \Omega)
\end{align}
(The $\delta(E_s + E_{s'} + \Omega)$ term was dropped because,
for $\Omega > 0$, the sum of the two positive energies $E_s$ and
$E_{s'}$ will never be negative.) The presence of the overall factor
$v_0$ in \eqref{movingframestimeintegral} and the absence of any
dependence on the trajectory in the volume integral in
\eqref{movingframesintegral} tells us that the perturbative scheme
in this section is non-relativistic.

We will proceed to solve for the complete set of modes,
$\chi_{(s|\pm)}({\bf x}) e^{\mp iE_s t}$, evaluating
\eqref{movingframesintegral} for necessary combinations of $s$ and
$s'$, before summing the squares of the resulting amplitudes to obtain
the pair production rates \eqref{movingframespairproductionrate}
and \eqref{movingframespairproductionrate_smallABphase}.

\subsection{Derivation of Equation \eqref{movingframesintegral}}
\label{movfraint}

We expand the Dirac operator, within the Heisenberg picture,
in terms of $\chi_{(s|\pm)}(x-\xi,y,z) e^{\mp iE_s t}$
\begin{align*}
\psi(t,{\bf x}) &= \sum_s \big[ \alpha_s(t) \chi_{(s|+)}(x-\xi,y,z)
                      e^{-iE_s t} \\
&\qquad + \beta_s^\dagger(t) \chi_{(s|-)}(x-\xi,y,z) e^{+iE_s t} \big]
\end{align*}
The $\alpha_s(t)$ and $\beta_s^\dagger(t)$ are time-dependent operators
that obey appropriate anti-commutation relations and ensure positivity
of the Hamiltonian. They also satisfy the equations of motion obtained
from the Heisenberg equations of motion:
\begin{align}
\label{heisenbergEOMs}
i\partial_t \psi &= h_\text{D} \psi \\
h_\text{D} &\equiv e A_0 -i \gamma^0 \gamma^j D_j + m \gamma^0 \nonumber
\end{align}

For a solenoid at rest at $\xi = 0$, the operators
\begin{align*}
a_s \equiv \alpha_s(t \to -\infty), \qquad
b_s \equiv \beta_s(t \to -\infty)
\end{align*}
acting on the zero particle state at $t \to -\infty$ destroy, respectively,
a fermion $(a_s)$ and an antifermion ($b_s$) associated with the
wavefunctions $\chi_{(s|\pm)}({\bf x})$. In particular,
\begin{align*}
\{a_s,a_{s'}^\dagger\} =
\delta_{s,s'}, \qquad \{b_s,b_{s'}^\dagger\} = \delta_{s,s'}
\end{align*}
Then, to leading order in the Hamiltonian, the zero particle to two
particle amplitude is
\begin{align}
&\label{movingframesamplitude}
{}_{\rm out} \langle s, s'| 0 \rangle_{\rm in} \nonumber \\
&\approx -i \left\langle 0; t = -\infty \left\vert a_s b_{s'}
\int dt d^3 x \ \psi^\dagger i \partial_t \psi
         \right\vert 0; t = -\infty \right\rangle
\end{align}
In \eqref{movingframesamplitude}, we have re-expressed the Dirac
Hamiltonian density
\begin{align*}
\mathcal{H} = \psi^\dagger h_\text{D} \psi
            = \psi^\dagger i\partial_t \psi
\end{align*}
using its Heisenberg equations of motion \eqref{heisenbergEOMs}.
(This is legitimate because the $\psi$'s appearing in the amplitude
\eqref{movingframesamplitude} are operator solutions to the Dirac
equation.) The reason for replacing $h_\text{D}$ with $i\partial_t$ is
the following. Since the time evolution of the $\alpha_s(t)$ and
$\beta_s(t)$ away from $a_s$ and $b_s$ are determined by the
interaction between $\psi$ and $A_\mu$, to lowest order in interaction,
we may now approximate
\begin{align*}
\alpha_s(t) \approx a_s, \qquad \beta_s(t) \approx b_s
\end{align*}
so that
\begin{align*}
\{a_s, \psi^\dagger\} &= \chi_{(s|+)}(x-\xi,y,z) e^{-iE_s t} \\
\{b_s, \psi\} &= \chi_{(s|-)}(x-\xi,y,z) e^{+iE_s t}
\end{align*}
By anticommuting the $a_s$ and $b_{s'}$ in
\eqref{movingframesamplitude} to the right, and noting that --
because the solenoid is moving solely in the $x$-direction --
\begin{align*}
\partial_t \chi(x-\xi(t),y,z) = -\dot{\xi}(t)
                                 \partial_x \chi(x-\xi(t),y,z),
\end{align*}
we then arrive at \eqref{movingframesintegral} after shifting integration
variables $x \to x-\xi$.

\subsection{Modes with $k_z=0$}

The solution for the gauge field of an infinite straight solenoid
along the $z$-axis with $\sigma=z$ has only two non-zero components
\begin{align}
\label{Fmunusolution}
F_{0y} = E^y &= \Phi \dot{\xi} ~ \delta(x-\xi(t)) \delta(y),
\nonumber \\
F_{xy} = -B^z &= -\Phi ~ \delta(x-\xi(t)) \delta(y),
\end{align}
The gauge potential that yields \eqref{Fmunusolution} is, in
Cartesian coordinates,
\begin{align}
\label{Amusolution_axial}
A_\mu = \left( 0,0,-\Phi \Theta(x-\xi(t))\delta(y),0 \right)
\end{align}
We wish to solve for the fermionic modes, $\chi$, in the stationary
solenoid ($\xi (t)=0$) case. It is easier to do so using the Lorenz
gauge, where in cylindrical coordinates,
\begin{align}
\label{Amusolution_lorenz}
A_\theta^\text{(S)} = \frac{\Phi}{2\pi}, ~
A_0^\text{(S)} = A_r^\text{(S)} = A_z^\text{(S)} = 0.
\end{align}
where the superscript $S$ denotes ``stationary''.

With the gauge potential determined, the Dirac equation now
reads\footnote{In this section, the mass term in the Dirac equation
has a $+$ sign, as opposed to the $-$ sign implied by
\eqref{diracaction}. To convert solutions for one into solutions
for the other, multiply the Dirac spinor by $\gamma_5$. Since this
corresponds to a change of basis, this choice of sign for the mass
term does not affect the results for the inner products.}
\begin{align}
\label{stationarystringdiraceqn}
\left[ i\gamma^\mu(\partial_\mu + ie A_\mu^\text{(S)})
        + m \right] \chi e^{-iEt} = 0
\end{align}
Translational symmetry along $z$ implies that
$\chi$ has the form
\begin{align}
\label{exponential}
\chi({\bf x}) = \varrho(r,\theta) e^{ik_z z}
\end{align}

Following Alford and Wilczek \cite{AlfordWilczek}, to exploit the
$z$ translational symmetry, it helps to find a set of
$\{\gamma^\mu\}$ matrices such that all of them except $\gamma^3$
are block diagonal. This way, in the reference frame where there is
no momentum along $z$, i.e. $k_z = 0$, the Dirac equation splits into
a pair of coupled equations, each involving only 2 component spinors.
The set of $\{\gamma^\mu\}$ we will use here are defined relative to
the ones in the chiral basis $\{\gamma_c^\mu\}$ in \eqref{chiralbasis}
as $\gamma^\mu \equiv U \gamma^\mu_c U^\dagger$ with
\begin{equation}
U \equiv \frac{1}{\sqrt{2}} \left[
\begin{array}{cccc}
0   & -i    & 0     & i     \\
-1  & 0     & -1    & 0     \\
0   & -1    & 0     & -1    \\
i   & 0     & -i    & 0
\end{array} \right]
\nonumber
\end{equation}
The new $\gamma^\mu$ are
\begin{align*}
\gamma^0 &= \left[ \begin{array}{cc}
-\sigma^3   & 0         \\
0           & \sigma^3
\end{array} \right], \qquad
\gamma^1 = \left[ \begin{array}{cc}
i \sigma^1  & 0         \\
0           & -i \sigma^1
\end{array} \right]     \\
\gamma^2 &= \left[ \begin{array}{cc}
-i \sigma^2 & 0         \\
0           & i \sigma^2
\end{array} \right], \qquad
\gamma^3 = \left[ \begin{array}{cc}
0               & -i \mathbb{I} \\
-i \mathbb{I}   & 0
\end{array} \right]
\end{align*}

Now the $k_z = 0$ solutions are
\begin{align}
\label{kz_0_solutions}
\chi_k(t,r,\theta,z) =
\left[ \begin{array}{c} \chi_A \\ \chi_B \end{array} \right]
e^{-i\omega_\pm t}, \ \ \omega_\pm \equiv \pm \sqrt{k^2+m^2},
\end{align}
with
\begin{eqnarray}
\chi_{\text{A}}^{(\sigma_1)} &=&
\mathcal{N}_{\text{A}} e^{in_1\theta} \left[
\begin{array}{c} J_{\sigma_1 (n_1+1+\epsilon)}(kr) e^{i\theta} \\
\frac{\sigma_1 k}{m+\omega_\pm} J_{\sigma_1 (n_1+\epsilon)}(kr)
\end{array} \right], \nonumber \\
\chi_{\text{B}}^{(\sigma_2)} &=& \mathcal{N}_{\text{B}}
e^{in_2\theta}
\left[
\begin{array}{c} \frac{\sigma_2 k}{m+\omega_\pm}
J_{\sigma_2 (n_2+1+\epsilon)}(kr) e^{i\theta} \\
J_{\sigma_2 (n_2+\epsilon)}(kr)
\end{array} \right],
\end{eqnarray}
where $\sigma_{1,2} = {\rm sgn}(n_{1,2}+\epsilon)$,
$n_{1,2} \in {\mathbb Z}$, $\epsilon =e\Phi/2\pi$ and the normalization factors $\mathcal{N}_{\text{A},\text{B}} \in \mathbb{C}$.
The spinors $\chi_\text{A}$ and $\chi_\text{B}$ are the two independent
solutions for each set of positive or negative energy states and
we have chosen our $z$-axis so that $\epsilon = e\Phi/(2\pi) > 0$.

For $n_{1,2} + \epsilon > 0$ and $n_{1,2} + \epsilon < -1$, the signs
$\sigma_{1,2}$ in the indices of the Bessel functions are fixed by
the requirement that the solutions $\chi_k$ must be square normalizable.
Specifically, the radial integral must converge at its lower limit.
It contains a factor of $r$, i.e. $\int_0 dr r$, whereas the Bessel
functions behave, for $kr \ll 1$, as $J_\nu \propto r^\nu$
(see \eqref{besselsmallz}).

When $n_{1,2}+\epsilon \in (-1,0)$, we have chosen $\sigma_{1,2} = -1$
so that the solutions $\chi_k$ remain square normalizable for all values
of magnetic flux $\Phi$. Even though both signs are allowed by square
normalizability for $\Phi \neq 0$, our choice of sign gives solutions
that, in the zero flux limit, join smoothly onto the cylindrical wave
solutions to the non-interacting massive Dirac equation
$(i\slashed{\partial}+m)\psi = 0$.

It is worth expanding upon this ambiguity and the resolution we have adopted
here. This is a subtle issue discussed first in context of strings in
Ref.~\cite{gerbert} (see also \cite{alfordnuclb}, \cite{jackiwgraphene}
and \cite{stone}). The basic problem is that for certain angular
momentum channels there are four solutions to the radial equation that
are normalizable rather than two (for a given energy and z-momentum). Thus
additional boundary conditions must be specified at the origin to determine
the two physical solutions. The set of possible boundary conditions is
restricted by the condition that only those solutions are permissible that
have zero radial current at the origin (``self-adjoint boundary
conditions''). However this does not by itself fully specify the boundary
conditions and some additional physical principle or regulation scheme must
be invoked. In effect we have made a special choice of boundary conditions
above. Our choice is natural and well-motivated for the following reasons:
(1) It extrapolates smoothly to the case of integer flux. (2) The transition
rate calculated using this boundary condition agrees with the perturbative
result in their common domain of validity
(see Sec.~\ref{movingframesrate} below). (3) One can
imagine natural regulation schemes in which this boundary condition will
arise \cite{alfordnuclb}. Still it must be kept in mind that the results
would come out different if different boundary conditions were used.

Shifting $n_{1,2}$ in \eqref{kz_0_solutions} by an integer,
$n_{1,2} \to n_{1,2} + m$ with $m \in \mathbb{Z}$, takes us
from one solution to another. Hence we can absorb the integer
part of $\epsilon$ in the label $n_{1,2}$ and, without loss
of generality, choose the AB phase $\epsilon$ to lie
in the interval $[0,1)$. We will denote the fractional part
of $\epsilon$ as $\kappa$, defined by
\begin{equation*}
\kappa = \epsilon\ {\rm mod}\ 1
\end{equation*}
In terms of $\kappa$, note that the set of mode functions here
(and the set with $k_z \neq 0$ below) is now manifestly periodic
in the AB phase $e\Phi$. Hence the radiation rate that
follow will also enjoy this periodicity.

\subsection{Modes with arbitrary $k_z$}

Above we have found all modes with $k_z=0$. Now we perform
boosts along the solenoid to obtain modes with $k_z \neq 0$.
Denote the Lorentz boost by $\mathcal{B}$ and so
\begin{align*}
\chi_k \to \mathcal{B} \cdot \chi_k
\end{align*}
Under the transformation
\begin{align*}
t &\to t \cosh\eta - z \sinh\eta
\end{align*}
with the rapidity parameter $\eta$ defined through
\begin{equation}
{\rm tanh}\eta = \frac{k_z}{k^0}
\end{equation}
the boost matrix $\mathcal{B}$ is
\begin{align*}
\mathcal{B} &= U \cdot \mathcal{B}_c \cdot U^{-1}, \\
\mathcal{B}_c &\equiv
\left[ \begin{array}{cc}
e^{\eta/2} \frac{1-\sigma^3}{2} + e^{-\eta/2} \frac{1+\sigma^3}{2} & 0 \\
0 & e^{\eta/2} \frac{1+\sigma^3}{2} + e^{-\eta/2} \frac{1-\sigma^3}{2}
\end{array} \right]
\end{align*}

Denote $\chi_\text{I}$ as the boosted $\chi_k$ with
$\mathcal{N}_\text{B} = 0$; and $\chi_\text{II}$ as the boosted
$\chi_k$ with $\mathcal{N}_\text{A} = 0$. The two independent
solutions for each set of positive and negative energy wavefunctions,
in the Lorenz gauge for $A_\mu$ \eqref{Amusolution_lorenz}, are then
\begin{align}
\label{lorenzgaugesolutions}
& \chi_{(\text{I},n,k,k_z|\pm)}(t,r,\theta,z) \nonumber  \\
&= \left[ \begin{array}{c}
\cosh(\eta/2) J_{\sigma(\kappa+n+1)}(kr) e^{i\theta} \\
\cosh(\eta/2) \sigma \lambda_\pm(k) J_{\sigma(\kappa+n)}(kr)  \\
-i \sinh(\eta/2) J_{\sigma(\kappa+n+1)}(kr) e^{i\theta} \\
i \sinh(\eta/2) \sigma \lambda_\pm(k) J_{\sigma(\kappa+n)}(kr)
\end{array} \right] \nonumber \\
& \qquad \times
\mathcal{N}_\pm(k,k_z) e^{\mp i k_0 t} e^{in\theta} e^{\pm ik_z z}, \\
& \chi_{(\text{II},n',k',k_z'|\pm)}(t,r,\theta,z) \nonumber \\
&= \left[ \begin{array}{c}
i \sinh(\eta'/2) \sigma' \lambda_\pm(k') J_{\sigma'(\kappa+n'+1)}(k'r)
             e^{i\theta} \\
-i \sinh(\eta'/2) J_{\sigma'(\kappa+n')}(k'r)  \\
\cosh(\eta'/2) \sigma' \lambda_\pm(k')
             J_{\sigma'(\kappa+n'+1)}(k'r) e^{i\theta} \\
\cosh(\eta'/2) J_{\sigma'(\kappa+n')}(k'r)
\end{array} \right] \nonumber \\
&\qquad \times \mathcal{N}_\pm(k',k_z')
e^{\mp i {k'}_0 t} e^{in'\theta} e^{\pm ik_z' z}, \nonumber
\end{align}
now with
\begin{align*}
\sigma, \sigma' = \left\{
\begin{array}{ll}
1   & \text{if }n,n' \geq 0   \\
-1  & \text{if }n,n' \leq -1
\end{array}\right.
\end{align*}
and
\begin{align}
\lambda_\pm(k) \equiv \frac{k}{m+\omega_\pm}, \ \
\mathcal{N}_\pm(k,k_z) \equiv
\sqrt{ \frac{\omega_\pm + m}{2 k_0 }}
\end{align}
\begin{align*}
k,k'>0; \ \ k_z,k_z' \in\mathbb{R}; \ \
n,n' = 0,\pm 1, \pm 2, \dots
\end{align*}

We have normalized our solutions using
\begin{align*}
\int_0^\infty dr ~ r ~
J_\nu(kr) J_\nu(k'r) &= \frac{\delta(k-k')}{\sqrt{kk'}}, \\
&\qquad k,k'>0, \ \text{Re}(\nu) > -1 \nonumber \\
\int_0^{2\pi} d\theta e^{i(n-n')\theta} &=
2\pi \delta_{n,n'}, \quad n,n' \in\mathbb{Z} \\
\int_{-\infty}^{+\infty} dz e^{i(q-q')z} &=
2\pi \delta(q-q'), \quad q,q' \in\mathbb{R}
\end{align*}
such that
\begin{align}
\label{normalization}
&\int d^3 x' ~
\chi^\dagger_{(A,n,k,k_z|\sigma_E)}({\bf x}')
\chi_{(A',n',k',k_z'|\sigma'_E)}({\bf x}') \nonumber \\
&\qquad = \delta_{\sigma_E,\sigma'_E} \delta_{A,A'} \delta_{n,n'}
(2\pi)^2 \frac{\delta(k-k') \delta(k_z-k_z')}{\sqrt{kk'}}, \\
&\qquad \sigma_E,\sigma'_E = \pm, \ A,A'\in\{\text{I},\text{II}\} \nonumber
\end{align}

We will be comparing the results here to those in
Sec.~\ref{infinitestraightstring}, where perturbation theory was
performed with plane wave solutions written in Cartesian coordinates. To make the comparison, it is worthwhile to note that the
normalization in \eqref{normalization} is consistent with the $\langle {\bf k}|{\bf k}' \rangle
= (2\pi)^3 \delta^3({\bf k} - {\bf k}')$ one would otherwise have
obtained if Cartesian coordinates were utilized. Moreover,
\begin{align*}
(2\pi)^3 &\delta^3({\bf k} - {\bf k}') \\
&= (2\pi)^3 \delta(\phi-\phi') \delta(k-k') \delta(k_z - k'_z)
\left(kk'\right)^{-1/2}.
\end{align*}
The $\sqrt{kk'}$ in \eqref{normalization} is the
(symmetrized) Jacobian when transforming from Cartesian to
cylindrical coordinates. The $(2\pi)\delta(\phi-\phi')$ is the
completeness relation
\begin{align}
\label{phicompleteness}
2\pi\delta(\phi-\phi') &= \sum_{n = -\infty}^{+\infty} e^{in (\phi-\phi')}, \quad |\phi-\phi'| \in [0,2\pi),
\end{align}
Denoting our mode functions here by $|n,k,k_z\rangle$
(and suppressing the $A$ and $\sigma_E$ dependence),
we see that \eqref{phicompleteness} accounts for the missing $(2\pi)$
in \eqref{normalization} because $|n,k,k_z\rangle$ is the $n$th term in
the Fourier series expansion of $|{\bf k}\rangle$, the mode functions
written in Cartesian coordinates, since
\begin{align*}
|{\bf k} \rangle =
| k, k_z, \phi \rangle = \sum_n | k, k_z, n \rangle e^{in\phi}.
\end{align*}
and hence \eqref{normalization} yields
\begin{align*}
\sum_{n,n'} \langle k,k_z,n | k', k'_z, n' \rangle
 e^{in'\phi'} e^{-in\phi} = \langle {\bf k} | {\bf k}' \rangle.
\end{align*}

\subsection{Transformation to Axial Gauge}

The above modes for a stationary solenoid, $\xi=0$, have been
found in Lorenz gauge. However, to find overlaps and the radiation
rate, it is easier to work
in axial gauge. Hence we now transform the modes to axial
gauge with $A_\mu$ in \eqref{Amusolution_axial},
\begin{align*}
A_\mu = \left( 0,0,-\Phi \Theta(x) \delta(y),0 \right),
\qquad \text{(Cartesian)}
\end{align*}
The $\chi$ solutions differ from its Lorenz gauge counterpart in
\eqref{lorenzgaugesolutions} only by a phase factor
\begin{align}
\label{axialgaugesolutions} \chi_\text{(axial)} =
e^{i\kappa(\theta \text{ mod } 2\pi)}
\chi_\text{(Lorenz, Eq \eqref{lorenzgaugesolutions})}
\end{align}

In \eqref{axialgaugesolutions}, to see it is $\kappa$ and not
$\epsilon$ that should occur in the phase, refer to the solutions
\eqref{kz_0_solutions} before the introduction of $\kappa$, and
note that the axial gauge version of \eqref{kz_0_solutions} would
contain a $e^{i\epsilon(\theta \text{ mod }2\pi)}$. The
$e^{i\kappa(\theta \text{ mod }2\pi)}$ in
\eqref{axialgaugesolutions} would then follow from the re-definition
of the Fourier mode label $n$ that introduced $\kappa$.

\subsection{Matrix Elements for Moving Solenoid}

With the solutions \eqref{axialgaugesolutions} in hand, we are
now ready to evaluate the integral \eqref{movingframesintegral}.
It is only necessary to calculate \eqref{movingframesintegral}
for positive and negative solutions from \eqref{axialgaugesolutions}
with Fourier mode labels $(n, n')=(-1,0)$ and $(0,-1)$,
where the $n$ refers to the positive energy solution and $n'$
to the negative energy solution. For other values of $n$ and $n'$,
the factor of $\partial_x \chi_{(s'|-)}$ in Eq.~(\ref{movingframesintegral})
yields the difference of two negative energy solutions in
\eqref{axialgaugesolutions}, which by \eqref{normalization} has
zero overlap with all its positive energy counterparts.

For $n'=0$, $\partial_x \chi_{(s'|-)}^{(\sigma'=+1)}$ is a linear
combination of the $n'=-1$ and the $n'=+1$ negative energy solutions
in \eqref{axialgaugesolutions}, except the former has
$\sigma'=+1$. For $n'=-1$, $\partial_x \chi_{(s'|-)}^{(\sigma'=-1)}$
is a linear combination of the $n'=-2$ and $n'=0$ negative energy
solutions in \eqref{axialgaugesolutions}, with the latter's
$\sigma'=-1$.
In addition, the $n'=-1$ solution with $\sigma'=+1$ in
$\partial_x \chi_{(n'=0|-)}$ remains orthogonal to all positive
energy solutions except possibly for the $n=-1$ case; likewise the
$n'=0$ solution with $\sigma'=-1$ in $\partial_x \chi_{(n'=-1|-)}$
is orthogonal to all $\chi_{(s|+)}$ except perhaps $\chi_{(n=0|+)}$.

As we will see in a moment, these two overlap integrals,
\begin{align*}
\int d^3 x' \left [ {\chi^{(\sigma=-1)}_{(n=-1|+)}}\right ]^\dagger
             \chi^{(\sigma'=+1)}_{(n'=-1|-)}
\end{align*}
and
\begin{align*}
\int d^3 x' \left [{\chi^{(\sigma=+1)}_{(n=0|+)}}\right ]^\dagger
             \chi^{(\sigma'=-1)}_{(n'=0|-)}
\end{align*}
do yield non-zero answers and are therefore the only ones contributing
to the pair production rate.

Computing the volume integral in (\ref{movingframesintegral}) for
solutions with the Fourier modes given by $(n,n')=(-1,0)$ and
$(0,-1)$ requires the integral
\begin{eqnarray}
\int_0^\infty dr ~ r ~ J_\nu(qr) J_{-\nu}(kr) && \nonumber \\
&& \hskip -7 cm
= \text{Pr}\left( \frac{2\sin(\pi\nu)}{\pi(q^2-k^2)}
           \left( \frac{q}{k} \right)^\nu \right) +
     \frac{\cos(\pi\nu)}{\sqrt{qk}} \delta(k-q), \nonumber\\
&\ \ k,q > 0, \ \text{Re}[\nu]>-1
\label{JnuJminusnuIntegral}
\end{eqnarray}
where $\text{Pr}$ denotes the principal value.

Define
\begin{equation*}
\mu_{++} \equiv
\sqrt{\frac{1}{\omega_k}+\frac{1}{k_0}}
\sqrt{\frac{1}{\omega_{k'}}+\frac{1}{k'_0}} -
\sqrt{\frac{1}{\omega_k}-\frac{1}{k_0}}
\sqrt{\frac{1}{\omega_{k'}}-\frac{1}{k'_0}}
\end{equation*}
and
\begin{equation*}
\mu_{+-} \equiv
\sqrt{\frac{1}{\omega_k}+\frac{1}{k_0}}
\sqrt{\frac{1}{\omega_{k'}}-\frac{1}{k'_0}} -
\sqrt{\frac{1}{\omega_k}-\frac{1}{k_0}}
\sqrt{\frac{1}{\omega_{k'}}+\frac{1}{k'_0}}.
\end{equation*}
The result then reads
{\allowdisplaybreaks
\begin{widetext}
\begin{eqnarray}
\label{movingframeNegativeToPositiveResults}
\left(\int d^3 x' \chi^\dagger_{(A,n,k,k_z|+)}({\bf x}')
\partial_x \chi_{(A',n',k',k'_z|-)}({\bf x}')\right)_{k \neq k'}
=  - \frac{1}{2} \delta(k_z+k'_z) \sin(\pi\kappa) \sqrt{kk'} &&
\nonumber \\
&& \hskip -12 cm
\times \biggl [
\delta_{n,0} \delta_{n',-1}
\left(\frac{k}{k'}\right)^\kappa
\bigg\{
\frac{\mu_{++}}{\omega_k+\omega_{k'}}
          \left ( \delta^{\text{I}}_{A} \delta^{\text{I}}_{A'}
            k \sqrt{\frac{\omega_{k'}-m}{\omega_k+m}}
%%%% b \beta
            + \delta^{\text{II}}_{A} \delta^{\text{II}}_{A'}
            k' \sqrt{\frac{\omega_k+m}{\omega_{k'}-m}} ~
          \right ) \nonumber \\
&&
\hskip -7.5 cm
+ \frac{\mu_{+-} i{\rm sgn}(k_z')}
    {(\omega_k-\omega_{k'}) \sqrt{\omega_k+m}\sqrt{\omega_{k'}-m}}
 \biggl ( \delta^{\text{I}}_{A} \delta^{\text{II}}_{A'} k k'
   - \delta^{\text{II}}_{A} \delta^{\text{I}}_{A'}
       (\omega_k+m)(\omega_{k'}-m)
  \biggr )
\biggr\}
\nonumber \\
&&
\hskip -12 cm
- \delta_{n,-1} \delta_{n',0} \left(\frac{k'}{k}\right)^{\kappa}
\bigg\{
\frac{\mu_{++}}{\omega_k+\omega_{k'}}
          \left ( \delta^{\text{I}}_{A} \delta^{\text{I}}_{A'}
            k' \sqrt{\frac{\omega_k+m}{\omega_{k'}-m}}
%%%% b \beta
            + \delta^{\text{II}}_{A} \delta^{\text{II}}_{A'}
            k \sqrt{\frac{\omega_{k'}-m}{\omega_k+m}} ~
          \right ) \nonumber \\
&&
\hskip -7.5 cm
+ \frac{\mu_{+-} i{\rm sgn}(k_z')}
    {(\omega_k-\omega_{k'}) \sqrt{\omega_k+m}\sqrt{\omega_{k'}-m}}
 \biggl ( \delta^{\text{I}}_{A} \delta^{\text{II}}_{A'}
                         (\omega_k+m)(\omega_{k'}-m)
   - \delta^{\text{II}}_{A} \delta^{\text{I}}_{A'} k k'
  \biggr )
\biggr\}
\biggr ]
\end{eqnarray}
\end{widetext}}

Direct substitution of the Bessel integral
Eq.~\eqref{JnuJminusnuIntegral} would lead to an additional
term in the transition amplitude above that is proportional to
$\delta ( k - k' )$. We have omitted this term following the discussion
in Ref.~\cite{AB_Bosons}. As explained there, the integrals must be
carefully regulated in order to exclude spurious terms that lead to
particle production even in the limit of zero flux. This is accomplished
by performing the integrals over a finite volume, imposing suitable
boundary conditions, and then taking the infinite volume limit. Such
an analysis eliminates the term proportional to $\delta( k - k' )$
(see Sec.~IIIB of ref \cite{AB_Bosons} for more details).

The expression in \eqref{movingframeNegativeToPositiveResults} yields
8 different channels for pair production to occur. Moreover, at the
level of individual amplitudes, the periodic dependence on the magnetic
flux is already manifest, since they depend on the AB phase only via
$\kappa = \epsilon \text{ mod } 1$.

\subsection{Moving frames pair production rate}
\label{movingframesrate}

Inserting \eqref{movingframeNegativeToPositiveResults} into
\eqref{movingframesintegral} and summing the squares of the
resulting individual amplitudes then gives us the rate of
fermion-antifermion pair production per unit length of the
infinite solenoid
{\allowdisplaybreaks
\begin{align}
\label{movingframespairproductionrate}
{\dot N}' &=
\int_0^\infty dk k \int_0^\infty dk'k'\int_{-\infty}^{\infty}dk_z
       \int_{-\infty}^{\infty} dk'_z \nonumber \\
&\qquad \times \delta\left( k_0 + k'_0  - \Omega \right)
   \delta(k_z+k'_z) \nonumber \\
&\qquad \times \frac{v_0^2 \sin^2\left( \pi \kappa \right)}
                    {8\pi^2\Omega^2 k_0 k'_0 }
       \left( m^2 + k_z^2 + k_0  k'_0  \right) \nonumber \\
&\qquad \times \left[
  \left( \frac{k}{k'} \right)^{2\kappa} + \left( \frac{k'}{k}
             \right)^{2\kappa} \right]
\end{align}}

In the limit when $\kappa \ll 1$, the last term (in parenthesis)
goes to $2$ and $\sin^2(\pi\kappa) \approx \pi^2 \kappa^2$, and
{\allowdisplaybreaks
\begin{align}
\label{movingframespairproductionrate_smallABphase}
{\dot N}' &=
\frac{v_0^2 \kappa^2}{4\Omega^2}
\int_0^\infty dk k\int_0^\infty dk'k'
   \int_{-\infty}^{\infty}dk_z \int_{-\infty}^{\infty} dk'_z \nonumber \\
&\qquad \times \delta\left( k_0  + k'_0  - \Omega \right) \delta(k_z+k'_z)
\left( 1+\frac{k_z^2+m^2}{k_0 k'_0} \right)
\end{align}}
This agrees with \eqref{infinitestraightstring_pairproductionrate_NR},
the non-relativistic limit of the interaction picture perturbation theory
result, if we identify $\kappa \leftrightarrow \epsilon$.

\section{Conclusions}
\label{conclusions}

We have solved for fermionic radiation first from oscillating
electromagnetic solenoids and then from cosmic string loops.
For the solenoid we have done the calculation in two different
ways, first using a small AB phase approximation, and second
by considering slowly moving solenoids. We have evaluated
the angular distribution of the fermionic radiation from the
solenoid, and the total power emitted from cosmic string
loops and cusps. Our results can be compared to the results
of Ref.~\cite{AB_Bosons}.

The total power emitted in bosons and fermions is very
comparable. For example, both are proportional to
$\epsilon^2 v_0^2 \Omega^2$ for the lowest harmonic of
the oscillating solenoid (see Eq.~\eqref{pprodrateleq1}).
However, the angular distributions of the radiation in
the two cases are quite distinct. To highlight the difference,
we show the angular distribution in both cases for the lowest
harmonic emission from an oscillating solenoid in
Fig.~\ref{angdistn}. We also find that the fermion
and antifermion are preferably emitted in opposite
helicity states and discuss the spin distribution.
Just like in the bosonic case, fermionic AB radiation
from kinks and cusps on cosmic strings is ultra-violet
divergent for massless fermions, with a linear
dependence on the cut-off. This may translate into
a significant amount of radiation of neutrinos from
strings with which neutrinos have an AB interaction.

Our results ought to apply to the low energy end of the emission
spectrum of electrons from idealized solenoids, where the wavelengths of the particle pairs are much longer than the diameter of the
solenoid. A more realistic theoretical investigation would have to
take into account the finite width of the solenoid itself.

In Ref.~\cite{AB_Bosons}, the gravitational analog of
AB radiation was also discussed. Via the same analogy,
we also expect cosmic strings to radiate fermions. We
leave that calculation for future work.

\begin{acknowledgments}
We thank Roman Jackiw for comments.
This work was supported by the U.S. Department of Energy at
Case Western Reserve University. TV was also supported by grant number
DE-FG02-90ER40542 at the Institute for Advanced Study. Much of the
numerical and analytic work in this paper was done with
{\sf Mathematica} \cite{mathematica}.
\end{acknowledgments}

\appendix

\section{Conventions}
\label{conventions}

Our metric and spacetime index convention are
defined by $\eta_{\mu\nu} = \text{diag}(1,-1,-1,-1)$,
$x^\mu = (t,x,y,z)$ with $\mu = 0,1,2,3$. Whenever $x$ appears
alone, without indices attached, it always means the $\mu=1$
component of $x^\mu$. The Einstein summation convention always
applies unless otherwise stated. The spacetime inner product of
$a^\mu$ and $b^\mu$ is $a \cdot b$; while $a^2 \equiv a_\mu a^\mu$.
Bold fonts denote spatial vectors; for instance,
${\bf a} \cdot {\bf b} = \delta_{ij} a^i b^j$ and
${\bf a}^2 = \delta_{ij} a^i a^j$.

{\it Dirac action:} The action in \eqref{diracaction} define the dynamics
of a Dirac fermion interacting with the photon vector potential $A_\mu$.
We make use of the Feynman slash notation. In particular,
\begin{align*}
\slashed{D} \psi &= \gamma^\mu(\partial_\mu + ie A_\mu)\psi
\end{align*}
Except in section \eqref{movingframesperturbationtheory}, we employ the
chiral basis for the $\{\gamma^\mu\}$ matrices,
\begin{align}
\label{chiralbasis}
\gamma^\mu &= \left[ \begin{array}{cc} 0 & \sigma^\mu \\
\bar{\sigma}^\mu & 0 \end{array} \right], \\
\sigma^\mu &= (\mathbf{1}_{2 \times 2}, \sigma^k), \quad
\bar{\sigma}^\mu = (\mathbf{1}_{2 \times 2}, -\sigma^k) \nonumber
\end{align}
which is, in turn, defined via the Pauli matrices,
\begin{align*}
\sigma^1 \equiv \left[ \begin{array}{cc} 0 & 1 \\ 1 & 0 \end{array} \right],
\quad
\sigma^2 \equiv \left[ \begin{array}{cc} 0 & -i \\ i & 0 \end{array}
\right],\quad
\sigma^3 \equiv \left[ \begin{array}{cc} 1 & 0 \\ 0 & -1 \end{array} \right]
\end{align*}

{\it Spinorial solutions:} Except in section
\ref{movingframesperturbationtheory}, interaction picture perturbation
theory will be used. It is carried out using the plane wave solutions to
the non-interacting massive Dirac equation
$(i\slashed{\partial}-m)\psi = 0$, where the positive energy
$\psi_{(k,s|+)}$ and negative energy $\psi_{(k,s|-)}$ solutions
with momentum $k$ and spin $s$ are
%$\psi_{(k',s|-)}$ solutions with
%momentum $k$ and $k'$ and spins $s$ and $s'$ are
\begin{align*}
\psi_{(k,s|+)} &\equiv
\frac{e^{-ik \cdot x}}{\sqrt{2k_0}} u_k^s
\equiv
\frac{e^{-ik \cdot x}}{\sqrt{2k_0}}
\left[
\begin{array}{c} \sqrt{\sigma \cdot k} ~\xi^s \\
\sqrt{\bar{\sigma} \cdot k} ~\xi^s
\end{array}
\right] , \\
\psi_{(k,s|-)} &\equiv
\frac{e^{+ik \cdot x}}{\sqrt{2k_0}} v_k^s
\equiv
\frac{e^{+ik \cdot x}}{\sqrt{2k_0}}
\left[ \begin{array}{c} \sqrt{\sigma \cdot k} ~\xi^{s} \\
- \sqrt{\bar{\sigma} \cdot k} ~\xi^{s} \end{array}
\right]
\label{eq:spinorsolution}
\end{align*}
where
\begin{equation}
(\xi^{s})_a = \delta^{s}_a,
\quad k_0 = \sqrt{{\bf k}^2+m^2}\ .
\end{equation}
The $\sqrt{\sigma \cdot k}$ and $\sqrt{\bar{\sigma} \cdot k}$ are
the matrices $\sigma \cdot p$ and $\bar{\sigma} \cdot p$ written in
diagonalized form ({\it i.e.} $U D U^{-1}$, where $D$ is diagonal),
with the eigenvalues replaced with their positive square roots.
By going to the rest frame of the particle, it can be seen that
the $\xi^1$ and $\xi^2$ are the spin up ($s = 1$) and spin down
($s = 2$) states for the fermion with respect to the basis of
Pauli matrices used here. For the antifermion, they are the spin
down ($s' = 1$) and spin up ($s' = 2$) states.
These plane wave solutions are normalized such that
\begin{align}
\label{planewavenormalization}
\int \psi^\dagger_{(k,s|\pm)} \psi_{(k',s'|\pm)} d^3 x =
(2\pi)^3 \delta^{(3)}({\bf k} - {\bf k}') \delta^s_{s'}
\end{align}

States of definite helicity are obtained by taking the spinors
$\xi$ to be eigenspinors of ${\boldsymbol \sigma} \cdot {\mathbf k}$
with eigenvalues $\pm | {\mathbf k} |$. For the particle the positive
eigenvalue corresponds to positive helicity and negative to negative;
for the antiparticle the negative eigenvalue corresponds to positive
helicity and vice-versa.

\end{document}